\documentclass[usenatbib,usegraphicx]{mn2e}
\usepackage{bm}

\title[The star cluster formation history of the LMC]
{The star cluster formation history of the LMC}
\author[Baumgardt et al.]{H. Baumgardt$^{1}$\thanks{E-mail:
h.baumgardt@uq.edu.au}, G. Parmentier$^{2,3}$, P. Anders$^{4}$ and E.~K. Grebel$^{5}$\\
$^{1}$School of Mathematics and Physics, University of Queensland, St. Lucia, QLD 4072, Australia \\
$^{2}$Max-Planck-Institut f\"ur Radioastronomie, Auf dem H\"ugel 69, D-53121 Bonn, Germany \\
$^{3}$Argelander-Institut f\"ur Astronomie, Universit\"at Bonn, Auf dem H\"ugel 71, D-53121 Bonn, Germany \\
$^{4}$Kavli Institute for Astronomy and Astrophysics, Peking University, Yi He Yuan Lu 5, Hai Dian District, Beijing 100871, China \\
$^{5}$ Astronomisches Rechen-Institut, Zentrum f\"ur Astronomie der Universit\"at Heidelberg, M\"onchhofstrasse 12~-~14, D-69120 Heidelberg, Germany
}

\begin{document}

\date{Accepted 2012 xx xx. Received 2012 xx xx; in original form 2012 xx xx}

\pagerange{\pageref{firstpage}--\pageref{lastpage}} \pubyear{201x}

\maketitle

\label{firstpage}

\begin{abstract}
The Large Magellanic Cloud is one of the nearest galaxies to us and is one of only few galaxies where the star formation history can be determined from studying resolved stellar populations. We have
compiled a new catalogue of ages, luminosities and masses of LMC star clusters and used it to determine the 
age distribution and dissolution rate of LMC star clusters. 
We find that the frequency of massive clusters with masses $M>5000$~M$_\odot$ is almost 
constant between 10 and 200 Myr, showing that the influence of residual gas expulsion is limited to the 
first 10 Myr of cluster evolution or clusters less massive than 5000~M$_\odot$.
Comparing the cluster frequency in that interval with the absolute star formation rate, we find that
about 15\% of all stars in the LMC were formed in long-lived star clusters that survive for more than 10 Myr.
We also find that the mass function of LMC clusters younger than $10^9$ Gyr can be fitted by a power-law mass function 
$N(m) \sim m^{-\alpha}$ with 
slope $\alpha=2.3$, while older clusters follow a significantly shallower slope and interpret this is a sign
of either incompleteness or the ongoing dissolution of low-mass clusters. Our data shows that for ages older than 200 Myr, about 90\% of 
all clusters are lost per dex of lifetime. The implied cluster
dissolution rate is significantly faster than that based on analytic estimates and $N$-body simulations.
Our cluster age data finally shows evidence for a burst in cluster formation about $10^9$ yrs ago,
but little evidence for bursts at other ages.
\end{abstract}


\section{Introduction} \label{sec:intro}

Open clusters are important test beds for star formation and stellar evolution theories since they
provide statistically significant samples of stars of known distance, age and metallicity.
Especially useful are star clusters in Local Group galaxies, which are close enough so that they 
can be resolved into individual stars and therefore allow their age and mass to be determined more accurately than based
on integrated photometry as is done for more distant clusters. 

The Large Magellanic Cloud (LMC) is one of the nearest galaxies to the Milky Way and contains several thousand star clusters \citep{betal08}
with ages ranging from a few Myr to a Hubble time. The large cluster system together with its proximity and the low extinction due to
the low inclination of the
LMC disc make the LMC cluster system an ideal test case to study the relation between star formation and cluster formation and 
the dissolution of star clusters.

Two main epochs of cluster formation have been found in the LMC, separated by an age
gap of several Gyr \citep{betal92, getal95, o96}. The first star formation episode led to the formation of globular clusters and ended
about 10 Gyr ago, while the second epoch started about 4 Gyr ago and is still ongoing. Within the last few Gyrs, several short peaks of
star formation have been found \citep[e.g.][]{hz09}, which might have been triggered by close encounters between the Large and Small Magellanic
Cloud or between the LMC and the Milky Way \citep{pu00, cetal06}. Similar peaks have also been found in the cluster formation rate and at least 
during the last Gyr, peaks in star cluster formation correspond to similar peaks that are seen in the field star formation rate \citep{mk11}.  

The mass function of star clusters in the LMC is generally found to follow a power-law $N(m) \sim M^{-\alpha}$, although the 
value of the power-law exponent $\alpha$ varies considerably between different authors.
\citet{cfw10} found that the mass function of LMC clusters more massive than $10^3$ M$_\odot$
can be fitted with a power-law slope $\alpha =1.8 \pm 0.2$ 
without any evidence for a change of this value with cluster age up to $10^9$ yrs.
de Grijs \& Anders (2006) found that cluster older than 100 Myr can be fitted by a power-law mass function with slope $\alpha= 2$, 
while younger clusters
follow significantly flatter mass functions. A slope between 2 and 2.4 was also found by \citet{hetal03} for clusters more massive
than a few times $10^3$ M$_\odot$ at $T=1$ Gyr and a few times $10^2$ M$_\odot$ at $T<10$ Myr in the combined 
cluster sample of the
LMC and SMC. On the other hand, from their reanalysis of the Hunter et al. cluster sample, Popescu et al. (2012) found a relatively small 
value of only $\alpha=1.5$ to 1.6 as the average slope for all clusters that are more massive than $10^3$ M$_\odot$ and less than a few Gyr old. 

Star clusters are prone to a number of dissolution mechanisms, including two-body relaxation and an external tidal field \citep{vh97, bm03}, 
disc and bulge shocking \citep{o72, go97} and encounters with giant molecular clouds \citep{s58, w85, getal06a}.
\citet{bl02} found evidence for ongoing dissolution of star clusters in the LMC with a characteristic lifetime of 
log($t^{dis}_4$/yr)=$9.7 \pm 0.3$ for a $10^4$ M$_\odot$ cluster, 
significantly longer than the corresponding value found by \citet{letal05b} for a sample of nearby galaxies. The long characteristic
timescale implies that the high mass end of the cluster mass function in the LMC has stayed virtually unchanged since the time of its formation.
Using the same data but a more careful treatment of the incompleteness limit, \citet{pg08} 
showed that the cluster disruption time scale is actually highly uncertain and only a lower limit of $t^{dis}_4 = 1$ Gyr can be given,
since clusters at the low-mass end fall below the completeness limit of \citet{hetal03} before they are destroyed.

Apart from the absolute time scale of cluster dissolution, another important question is whether the timescale for cluster
dissolution depends on the cluster mass.
Using the cluster sample of \citet{hetal03} and based on the observed relation between the mass of the most massive
cluster and the sampled age range, \citet{gb08a} found evidence for mass dependent cluster dissolution for ages larger than $10^8$ yrs,
but mass independent dissolution for smaller ages. \citet{cfw10} on the other hand found no evidence for mass dependent cluster dissolution up 
to $10^9$ yrs when looking at the evolution of the mass function of LMC and SMC star clusters.

Most of the above papers are based on data from the cluster catalogue of \citet{hetal03}, who determined properties of 939 clusters in the SMC 
and LMC
based on ground-based CCD photometry. Recently a wealth of new data has been obtained for LMC star clusters by \citet{ggk10} and 
\citet{phe12}. Glatt et al. have presented
ages and luminosities of almost 1200 star clusters in the LMC using data from the Magellanic Cloud Photometric Surveys \citep[MCPS]{z02, z04}. 
In addition, Popescu et al. (2012) have presented new age and mass estimates based on integrated colors for the clusters of \citet{hetal03} taking into account
random sampling of the IMF and stochastic effects due to bright stars. 

In this paper, we present a new derivation 
of the mass and age distribution of LMC clusters
and use it to study the relation between star and star cluster formation and different cluster dissolution models. 
Our paper is organised as follows: In Sec.~2 we present the observational data for LMC clusters and derive a combined
catalogue of ages and masses for massive LMC clusters. We use this catalogue in Sec.~3 to calculate the age distribution
and mass function of LMC clusters. In Sec.~4 we present a discussion of the various cluster 
dissolution mechanisms and compare predictions with the observational data in Sec.~5. Our conclusions are finally presented in Sec.~6.

\begin{figure}
\begin{center}
\includegraphics[width=84mm]{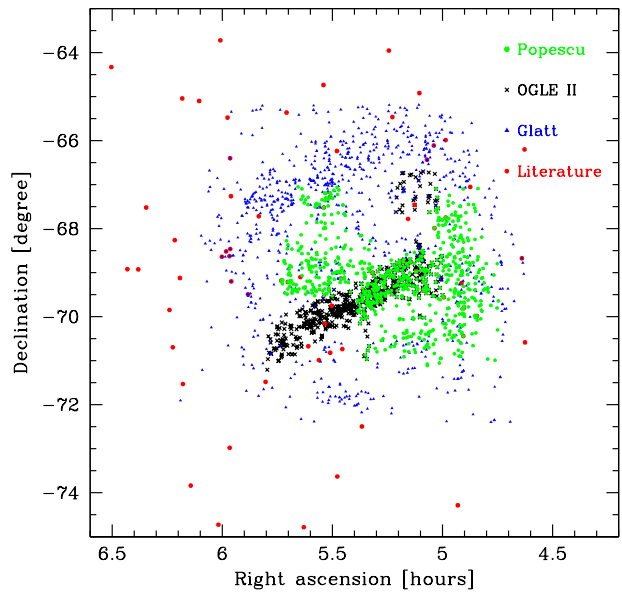}
\end{center}
\caption{Position of star clusters in the LMC from recent large surveys plus a few additional clusters from the literature. Only Glatt et al. (2010) and
\citet{mg03} cover the whole LMC, while all other surveys are restricted to certain parts of the LMC.}
\label{fig1}
\end{figure}

\section{Observational data} \label{sec:obs}

We use four recent compilations of LMC star cluster parameters to derive a 
combined catalogue for LMC clusters. Our first catalogue is the recent work of \citet{ggk10},
who used data from the Magellanic Cloud Photometric Surveys together with isochrone fitting to
derive ages and luminosities for 1193 populous star clusters in a 64 deg$^2$ area of the LMC. 
The cluster identifications, sizes, and positions were taken from the catalogue of \citet{betal08}.
Since the MCPS becomes highly incomplete below $V=24$, the Glatt et al. data become incomplete
for ages larger than 500 Myr and do not contain any clusters older than $T = 1$ Gyr. 
Glatt et al. also did not determine ages for clusters younger than about 10 Myr to avoid confusion between
star clusters and unbound stellar associations. 

The second data set that we use is the catalogue of \citet{pu00}, who used {\it BVI} data from the
OGLE II microlensing experiment \citep{u98} together with isochrone 
fitting to derive  ages of about 600 star clusters which are located in the central parts of the LMC
and are younger than 1.2 Gyr.
Since Pietrzynski \& Udalski did not derive cluster luminosities or cluster masses, we could only use 
their data if luminosity information was available from another source. Our third catalogue is the
cluster catalogue presented by \citet{hetal03}. The Hunter et al. catalogue contains data for about 746 different
LMC clusters, limited to the inner regions of the LMC and to absolute luminosities brighter than $M_V=-3.5$. 
New age determinations for the cluster in the Hunter et al. catalogue were presented by 
\citet{dga06} from broad-band spectral energy distribution fitting of the integrated UBVR photometry 
given by \citet{hetal03}
and recently also \citet{phe12}. Popescu et al. use the stellar cluster analysis package
{\tt MASSCLEANage} to derive new ages and masses for the Hunter et al. cluster sample. The advantage of 
{\tt MASSCLEANage} is that it allows to take into account stochastic fluctuations in cluster colors and magnitudes 
which arise from the finite number of
stars present in clusters. We average the ages for clusters which appear multiple times in the Popescu et al. 
sample and are left with a list of 746 unique clusters which are almost identical to those analysed 
by de Grijs \& Anders.

In order to increase the accuracy of the age and mass determinations especially for bright clusters, we also include 
age determinations from the literature. In particular, we use the ages recently derived by \citet{metal09}
for massive intermediate-age star clusters from colour-magnitude
fitting and the age data compiled by \citet{mg03} for 53 massive LMC clusters. Many of the ages in these papers are 
derived from {\it HST} data reaching several magnitudes below the turnover and are therefore significantly more accurate 
than the ages derived in the above surveys.


\begin{figure*}
\begin{center}
\includegraphics[width=160mm]{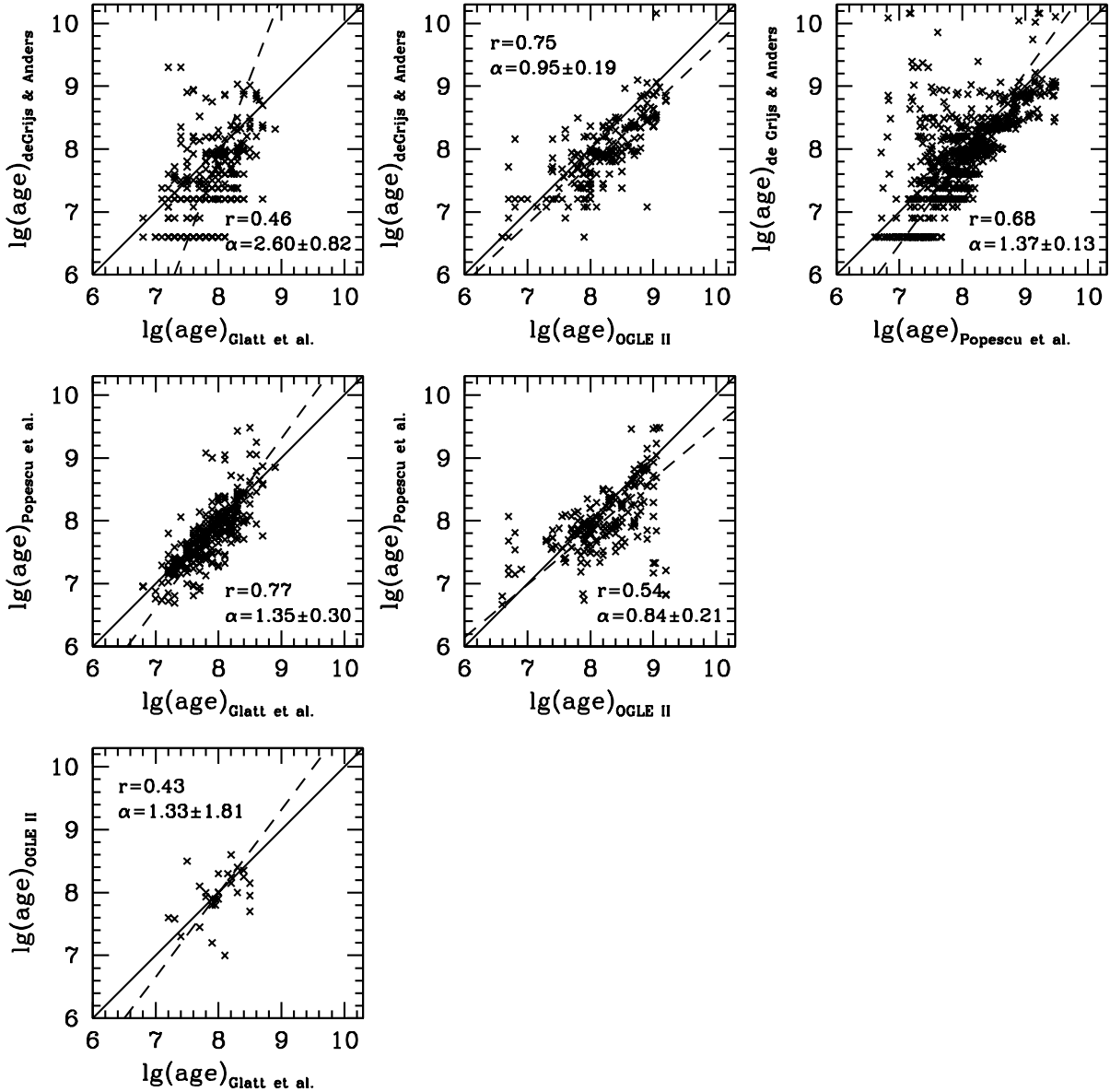}
\end{center}
\caption{Comparison of age determinations from Pietrzynski \& Udalski (2000) (OGLE II), de Grijs \& Anders (2006), Glatt et al. (2010)
and Popescu et al. (2012). Crosses denote the positions of individual clusters. Solid lines show lines of equal ages,
while dashed lines give the best-fitting linear relation. Shown are also the Pearson $r$ coefficients and the
slope $\alpha$ of the best fitting relation for the different comparisons.}
\label{fig2}
\end{figure*}

Fig.\ \ref{fig1} depicts the positions of all clusters from the above catalogues. Since the list of clusters in 
de Grijs \& Anders 
is nearly identical to that of Popescu et al., we do not show them separately. It can be seen that
the Glatt et al. data covers most of the LMC, while \citet{dga06}, \citet{phe12}
and \citet{pu00} catalogues are restricted to varying parts of the LMC. Since the LMC shows
evidence for different regions having undergone different star formation histories \citep{ggk10},
it is clear that only a combination of all catalogues can give a complete picture of the LMC cluster
population.

\begin{table*}
\caption{Number of clusters $N_{Cl}$ in common between different catalogues, Pearson $r$ coefficient, slope of
best-fitting relation $\alpha$ in an log(age) vs. log(age) diagram, and the residual scatter $\log T$ about this 
best-fitting relation for all data sets.}

\medskip

\begin{tabular}{l@{\,\,}l@{}r@{\,\,}c@{}c@{}c|l@{\,\,}l@{}r@{\,\,}c@{}c@{}c}
\hline
Data set 1 & Data set 2 & $N_{Cl}$ & Pearson & Slope & Scatter & Data set 1 & Data set 2 & $N_{Cl}$ & Pearson & Slope & Scatter \\
 & & & $r$ & $\alpha$ & $<\!\Delta \log T\!>$ & & & & $r$ & $\alpha$  & $<\!\Delta \log T\!>$\\
\hline
OGLE II & de Grijs\&Anders & 198 & 0.75 & $0.95 \pm 0.19$ & 0.28 & de Grijs\&Anders & OGLE II           & 198 & 0.75 & $1.05 \pm 0.21$ & 0.28 \\
OGLE II & Glatt et al.    & 29  & 0.44 & $0.75 \pm 1.38$ & 0.26 & de Grijs\&Anders & Glatt et al.      & 294 & 0.46 & $0.38 \pm 0.11$ & 0.33 \\
OGLE II & Popescu et al.  & 191 & 0.54 & $0.84 \pm 0.22$ & 0.37 & de Grijs\&Anders & Popescu et al.    & 745 & 0.68 & $0.73 \pm 0.07$ & 0.62 \\
OGLE II & Literature val. & 11  & 0.96 & $2.07 \pm 1.40$ & 0.13 & de Grijs\&Anders & Literature val.   &  18 & 0.91 & $1.60 \pm 0.67$ & 0.27 \\
 & & & \\
Glatt et al. & OGLE II            & 29   & 0.44 & $1.33 \pm 1.81$ & 0.26 & Popescu et al. & OGLE II           & 191 & 0.54 & $1.19 \pm 0.31$ & 0.37 \\
Glatt et al. & de Grijs\&Anders    & 294  & 0.46 & $2.61 \pm 0.83$ & 0.33 & Popescu et al. & Glatt et al.      & 293 & 0.77 & $0.74 \pm 0.16$ & 0.20 \\
Glatt et al. & Popescu et al.     & 293  & 0.77 & $1.35 \pm 0.30$ & 0.20 & Popescu et al. & de Grijs\&Anders   & 745 & 0.68 & $1.37 \pm 0.13$ & 0.62 \\
Glatt et al. & Literature val.    & 18   & 0.70 & $1.71 \pm 1.56$ & 0.30 & Popescu et al. & Literature val.   &  18 & 0.07 & $ 3.99 \pm 0.67$ & 0.82 \\
\hline
\end{tabular}
\end{table*}

Fig.\ 2 and Table~1 compare the ages of star clusters derived in the studies by Pietrzynski \& Udalski,
de Grijs \& Anders, Glatt et al., and Popescu et al. with one another for the clusters common to 
the different studies. The de Grijs \& Anders ages are in good agreement with the OGLE II ages
but disagree significantly with the Glatt et al. and Popescu et al. ages, since clusters with $T<10^8$ yrs in 
de Grijs \& Anders tend on average to be younger than found by the other authors, while clusters with $T>10^8$ yrs 
are on average older in de Grijs \& Anders than in the other catalogues.
As noted by Popescu et al., this difference could be due to the absence of single red giants or supergiants in 
young clusters, which make the cluster appear
bluer and therefore younger. The Glatt et al. data shows fairly good agreement with Popescu et al. It is also
in rough agreement with the OGLE II data, however the number of clusters in common is too small to make
a meaningful comparison. The OGLE II ages 
have a small offset by about $\log T = 0.2$ dex compared with the de Grijs \& Anders ages, which might also be present
in the comparison with the Popescu et al. data.
As noted by de Grijs \& Anders, this can be explained by the smaller LMC distance modulus of $(m-M)=18.23$ adopted by
Pietrzynski \& Udalski, compared to the value of $(m-M)=18.50$ used by most other authors. Placing
the LMC at a smaller distance decreases the absolute luminosities of the stars and the turn-over region in
the CMD, and therefore makes the clusters appear older. In order to correct for this bias, we increase the OGLE II
ages by 0.2 dex.

From the comparison of the cluster ages shown in Fig.\ \ref{fig2}, we calculate Pearson $r$ coefficients
and best-fitting linear relations $\log \mbox{age}_1 = \alpha \log \mbox{age}_2 + \beta$ and the standard deviation
in $\log T$ about this best-fitting relation.
The results are shown in Table~1, which also includes a comparison with the literature data.
The relatively large discrepancy between the de Grijs \& Anders data on the one hand and
Glatt et al. and Popescu et al. on the other is again apparent. Since Glatt et al. use CMD fitting while Popescu et al. work with 
integrated colors, the difference cannot result from the method used and is more likely due to stochastic effects
influencing the integrated colors. Given the Pearson $r$ coefficients, the ages from OGLE II seem to 
better agree with de Grijs \& Anders ages than with either Glatt et al. or Popescu et al.
However the slope $\alpha$ of the best-fitting relation for OGLE II is in all cases compatible with unity.
The literature sample has only few clusters in common with the other 
studies, so it is hard to make a meaningful comparison, but the Pearson $r$ coefficients as well as
the slopes $\alpha$ are mostly compatible with unity within the error bars, indicating good agreement. 
The final column of Table~1 shows that the RMS deviation in $\log t$ between the different catalogues 
is typically around 0.20 to 0.30, a value that also agrees with the age uncertainties of individual clusters
(see Table 2).

To obtain a complete catalogue of LMC clusters, we combine the clusters from the above data sets, 
excluding duplicate entries through a coordinate based search with a search radius of 20'', which corresponds
to a distance of 5~pc at an assumed LMC distance of 50 kpc. When
assigning ages to the clusters, we give the highest priority to the ages given in
\citet{metal09} and Mackey \& Gilmore (2003), followed by Glatt et al. (2010), the OGLE II data, and finally 
Popescu et al. (2012). We ignore the de Grijs \& Anders dataset due to its significant deviation against
the Glatt et al. and Popescu et al. ages. 
Since we need cluster luminosities 
to calculate cluster masses and OGLE II has only age information, we include clusters from OGLE II 
only if information on the total cluster luminosity is available from 
Popescu et al. (2012). 

Cluster masses are then derived from the $M_V$ luminosities by calculating
$M/L_V$ values using the Padova isochrones for $Z=0.008$.
These isochrones are based on the Marigo et al. (2008) stellar evolution tracks with corrections from case~A in 
Girardi et al. (2010), assuming a Kroupa (1998) stellar mass function corrected for binaries. 
The resulting cluster masses show good agreement with the cluster masses calculated by Popescu et al. for the clusters
which we take from their catalogue. 
As shown by Anders et al. (2009), cluster dissolution can affect the cluster $M/L$ ratios since low-mass 
stars are lost preferentially, however the resulting effect on the $M/L$ ratios is less than 50\% until
the cluster is close to dissolution.

\begin{figure}
\begin{center}
\includegraphics[width=84mm]{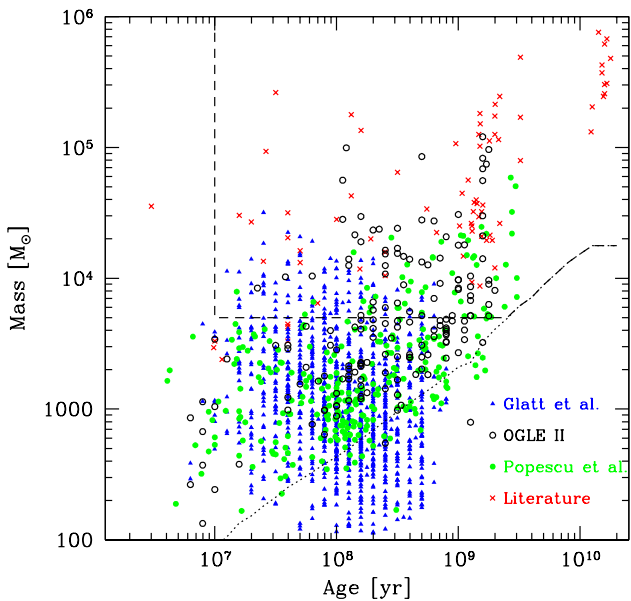}
\end{center}
\caption{Location of all clusters in a mass vs. age diagram. The dotted line shows the location of the 50\% completeness limit of $M_V=-3.5$
for the Popescu et al. (2012) data. The dashed line shows the location of clusters used for our analysis, which are
confined to ages larger than $10^7$ yrs, masses larger than 5000 M$_\odot$ and total luminosities brighter than $M_V=-3.5$.
}\label{fig3}
\end{figure}

Fig. \ref{fig3} depicts the location of all clusters from the combined data set in a  mass vs. age diagram.
The dotted line shows the 50\% completeness limit of $M_V=-3.5$ from \citet{hetal03}.
The majority of clusters have ages between $10^7$ to $10^9$ yrs and masses of a few hundred to a few thousand
M$_\odot$. The lack of clusters younger than $10^7$ yrs is due to the fact that such clusters are often obscured
by dust clouds and are also easy to
confuse with unbound associations and were therefore not analysed by Glatt et al. and OGLE II. The Glatt et al. 
and OGLE II samples also do not contain clusters older than $\approx 10^9$ yrs, since neither the OGLE nor the MCPS 
photometry is deep enough to permit age dating such clusters from resolved CMDs. This could
in part explain the absence of clusters above $10^9$ yrs. 
In addition, the 50\% completeness limit of the Hunter et al. sample reaches several thousand M$_\odot$ for ages
larger than $10^9~\mbox{yrs}$. In order to work with a sample that is as complete as possible, we therefore
restrict our analysis to clusters older than $10^7$ yrs and brighter than $M_V=-3.5$. For clusters
of a few thousand stars, a single giant star can have a luminosity exceeding that of the rest of the cluster,
leading to large fluctuations in the clusters' $M/L$ ratio \citep{petal11}.
This makes mass estimates based on cluster luminosity highly uncertain. 
We therefore also restrict our analysis to clusters more massive than 5000 M$_\odot$. For clusters of this mass
and ages larger than $10^7$ yrs, Fig.\ 11 in \citet{petal11} shows that stochastic effects due to bright 
giants change the $M/L$ ratio in the V band by less than 20\% compared to a model with a continuous IMF. Comparable
mass uncertainties are also found by Popescu et al. for clusters more massive than a few thousand solar masses. 
The resulting
selection limit is shown in Fig. \ref{fig3} by dashed lines. Out of a total number of 1649 unique clusters in 
the four data sets, 322 clusters pass our selection criteria. Their basic parameters (names, positions,
total $M_V$ luminosities, ages, age uncertainties and masses) are listed in Table 2. Although containing only 
the brightest and most massive clusters in the LMC, 
none of the individual catalogues is complete within our age and mass ranges: Of the 322 clusters in our final sample, only
86 were listed in OGLE II, 85 in Glatt et al. and 194 in Popescu et al. In the following analysis, we restrict ourselves
to the area of the LMC covered by Glatt et al., which is roughly located between 4.7 and 6.1 hours in right ascension
and -65 and \mbox{-72.5} degrees in declination. This region contains 294 clusters of our final sample and is identical
to the region for which \citet{hz09} determined the field star formation rate, which we will compare to the
cluster formation rate in the next
section.

\section{Results} \label{sec:results}

\begin{figure}
\begin{center}
\includegraphics[width=84mm]{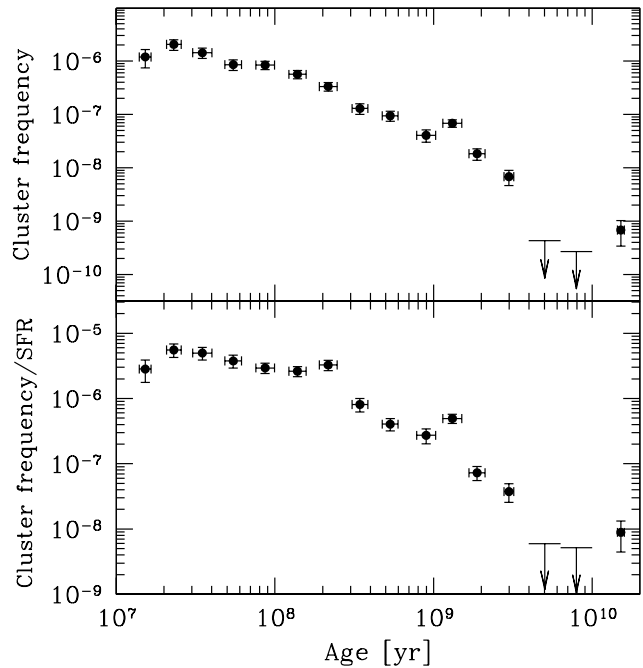}
\end{center}
\caption{Upper panel: Cluster frequency $dN/dt$ as a function of age. Solid points show the observed LMC data
for clusters with masses $M_C>5000$ M$_\odot$. Lower panel: Cluster frequency divided by the average star formation
rate in each age bin. The number of clusters formed per unit stellar mass is constant up to an age of 200 Myr and drops sharply
for larger ages, indicating the onset of cluster dissolution for ages $T>200$ Myr.}
\label{fig4}
\end{figure}

Fig.\ \ref{fig4} depicts the cluster frequency $dN/dt$, defined as the number of clusters per time unit, as a
function of age for clusters with present masses $M_C>5000$ M$_\odot$ and magnitudes brighter than $M_V=-3.5$. It can be seen that 
the observed frequency drops steadily
as a function of age. The drop is relatively small during the first 100 Myr since the data point at T=150 Myr is only a 
factor of two lower than the one at the youngest age. 
The data show clear evidence for a peak in cluster frequency at $T=1.5 \cdot 10^9$ yrs. Other peaks are more ambiguous,
but could be present between 20 to 30 Myr and around 100 Myr. The peak at ages between 20 to 30 Myr,
might however simply be due to cluster incompleteness at ages $T<20$ Myr. Also visible is the well-known absence of star clusters 
between 3 and 8 Gyr, where the cluster frequency is at least a factor of 10 lower than at older and younger ages. 

In order to disentangle true variations in cluster frequency from star formation variations, we divide the cluster frequency
by the total star formation rate in each age bin according to \citet{hz09} and show the result in the bottom panel of Fig. \ref{fig4}.
As can be seen, the cluster to star ratio is now nearly constant for ages from 10 Myr to 200 Myr. 
Primordial gas expulsion does therefore either have
only a moderate effect on star clusters or its observable effects must be limited to either the first 10 Myr of cluster     
evolution or low-mass clusters with $M_C<5000$ M$_\odot$ since no strong drop in cluster frequency is visible in the first 100 Myr.
After 200 Myr, the frequency of clusters drops significantly, 
at $T=1$ Gyr it is roughly only 1/10th of that at $T<200$ Myr, and at $T=10$ Gyr it is lower by an additional
factor of 10. We note that at $T=10$ Gyr, the 50\% completeness limit of
\citet{hetal03} is almost a factor of 4 more massive than our
mass-cut at 5000 M$_\odot$ (see Fig.~4).  This, however, does not explain the low number of clusters at $T=10$ Gyr,
since all of them are more massive than $10^5$ M$_\odot$. 

A nearly constant cluster frequency at young ages followed by a strong decrease at later ages was
also found by \citet{letal05a} for the open clusters in the solar neighborhood. In the solar neighborhood
the break occurs at around $\log{t/yr} =8.6$, equivalent to 400 Myr. It is also visible in the distribution
of massive star clusters with $M>10^3$ M$_\odot$ in the SMC resented by \citet{cfw06} (see their Fig. 1).
\citet{bl03} showed that this
behavior is due to cluster dissolution: For ages $t<t_{break}$, where $t_{break}$ is a characteristic time related to the dissolution
time, only a small number of low-mass clusters fall below the detection threshold, so the cluster frequency drops only slowly,
while for ages $t>t_{break}$ significant dissolution of clusters causes a strong decrease in cluster frequency. 

From Fig. \ref{fig4}, we can therefore conclude that a) if residual gas expulsion has an influence on star clusters,
its effects must be limited to either the first 10 Myr of evolution or low-mass clusters with $M_C<5000$ M$_\odot$ b) the characteristic lifetime of a 
$\sim 10^4$ M$_\odot$ star cluster (which make up the majority of our clusters in our sample) in the LMC is a few hundred Myr and roughly
the same as in the solar neighborhood \citep{letal05a}; c) for ages $>200$~Myr about 90\% of clusters are destroyed
per 1 dex in $\log t$. 

The data shown in Fig.\ \ref{fig4} confirm the absence of star clusters in the LMC in the age range 4 Gyr $< T < 10$~Gyr, which 
was first noticed by \citet{betal92}. A similar gap is not seen in the field star formation rate, as found by \citet{hz09}
from an analysis of the MCPS data and \citet{hetal99} from an analysis of {\it HST} color-magnitude diagrams of three fields in the LMC.
If the star formation rate derived by \citet{hz09} is correct, then the ratio of the number of clusters formed to 
the absolute star formation rate was at least a factor 5 to 10 lower in this age range then at both earlier ages $T>10$~Gyr and
later ages $T<3$ Gyr. Alternatively, only low-mass clusters with $M< 5000$ M$_\odot$ may have been formed in this age range.

\begin{figure}
\begin{center}
\includegraphics[width=84mm]{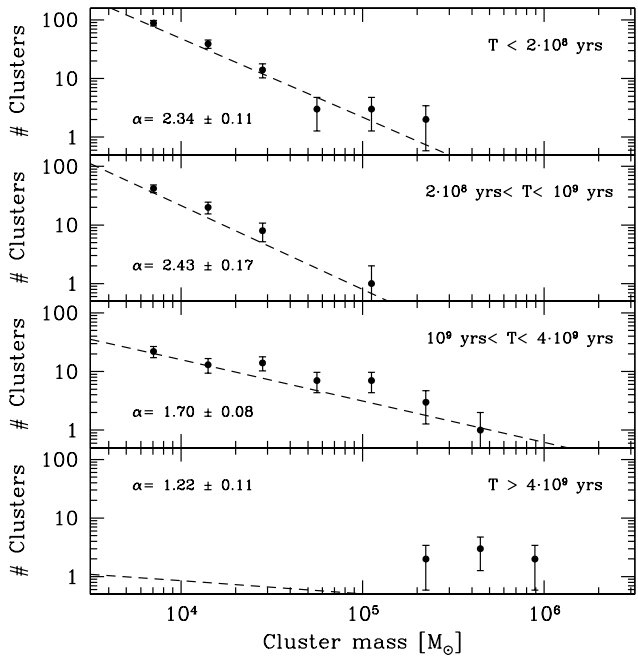}
\end{center}
\caption{Mass distribution of star clusters for four different age bins. The dashed lines show best-fitting power-law mass functions. 
The power-law slopes $\alpha$ and their errors are also given in the different panels. Clusters in the two youngest age bins follow 
power-law mass functions with slope $\alpha \approx 2.35$. Clusters older than 1 Gyr follow a significantly flatter distribution, 
which is most likely a result of ongoing cluster dissolution. For clusters with $T>4$ Gyr, a power-law mass function does not 
give a good description of their mass function.}
\label{fig5}
\end{figure}

Fig.\ \ref{fig5} finally depicts the mass distribution of clusters, split into four age groups. In most age bins, the mass function of clusters
can be approximated well by a power-law mass function $N(m) \sim m^{-\alpha}$. Applying a maximum likelihood estimator to the cluster mass distribution 
as described in \citet{crn09}, gives as best-fitting slope $\alpha=2.32 \pm 0.11$ for the youngest age bin. This is in good agreement with the slope found
by \citet{hetal03} and only slightly steeper than the slope of $\alpha=1.8 \pm 0.2$ found by \citet{cfw10} for clusters more massive than
$10^3$ M$_\odot$. 
In our sample, clusters up to 1 Gyr can be fitted by a slope similar to that for clusters in the youngest age bin, confirming results by
\citet{cfw10} that the dissolution of clusters up to 1 Gyr is mass independent, at least for the mass range considered here.
Clusters between 1 and 4 Gyr have a significantly flatter best-fitting slope of $\alpha=1.67 \pm 0.08$. This could be due
to either incompleteness of the sample at the low-mass end or be a sign of ongoing 
cluster dissolution, since low-mass clusters are more easily destroyed by two-body relaxation than high-mass ones \citep{bm03}.
If the flattening were due to incompleteness and the true mass function still a power-law with slope $\alpha=2.3$, then
about 400 clusters with masses between 5000~M$_\odot$ and $10^5$~M$_\odot$ would be missing from our sample. This seems highly 
unlikely, given the relatively faint 50\% completeness limit of the Hunter et al. sample (about 5000~M$_\odot$ at $T=2$~Gyr).
We therefore conclude that the observed flattening of the mass function for clusters with ages $1<T<4$ Gyr is 
due to ongoing cluster dissolution.
Interestingly, clusters with ages $1<T<4$ Gyr and masses larger than $10^5$~M$_\odot$ can still be fitted with a slope 
$\alpha \approx 2.3$, the slope flattens only for lower-mass clusters, which is also in agreement with ongoing 
dissolution of low-mass clusters (see sec. \ref{sec:cth}).
The mass-function slope is even flatter for clusters in the oldest age bin (bottom panel of Fig.\ \ref{fig5}), however for these a power-law mass 
function provides a rather poor fit and their mass function is much better described by e.g. a log-normal distribution, 
similar to what is seen for globular clusters in other galaxies.

From Figs. \ref{fig4} and \ref{fig5}, we can also deduce the ratio of stars born in clusters that survive gas expulsion and infant mortality 
to the total number of stars born.
Our sample contains 87 clusters with ages 10 Myr $< T < 100$ Myr and masses $M_C>5000$ M$_\odot$, which have a total mass of
$1.3 \cdot 10^6$ M$_\odot$. If we assume that clusters follow a power-law mass function with exponent $\alpha=2.3$
and extend the cluster mass function down to clusters of $10^2$ M$_\odot$, the total mass in clusters is about a factor 3
higher. 
Integrating the \citet{hz09} star formation rate, we find that over the same time span $2.6 \cdot 10^7$ M$_\odot$ were born in
field stars. Hence about 15\% of all stars with $T<100$ Myr formed in clusters that survive for at least 10 Myr. This is within  
the range of ratios seen in other nearby galaxies as determined by \citet{gbk10}.

\section{Theoretical estimates for the dissolution of star clusters}\label{sec:cldiss}

Star clusters are destroyed through a number of mechanisms,
including residual gas expulsion \citep{h80, g97, bk07}, two-body relaxation and external tidal fields
\citep{vh97,bm03} and tidal shocks from disc or bulge passages or passing giant molecular clouds 
\citep{s58,o72,w85,go97,getal06a}. In the following, we discuss each of these dissolution mechanisms in more detail.

\subsection{Gas expulsion}

Clusters typically form at the centres of dense molecular clouds with star formation efficiencies
of about 30\%. When the residual gas that is not converted into stars is removed by winds from young
O and B type stars, the gravitational potential is lowered and some clusters can become unbound.
\citet{bk07} showed that the impact of gas expulsion depends on the star formation efficiency
$\epsilon$, the timescale over which the gas is removed compared to the clusters crossing time
$\tau_{gas}/t_{cross}$, and the strength of the external tidal field which can be quantified through the ratio of the
clusters half-mass radius prior to gas expulsion to its tidal radius $r_h/r_t$. 
Although gas expulsion is thought to happen within the first few Myr after cluster formation, and therefore does not directly
affect clusters with ages $\ge 10$~Myr, clusters need several initial crossing 
times to dissolve into the field. While dissolving, they could still be visible as clusters and would therefore
influence our sample at the youngest ages (see e.g. Parmentier \& Baumgardt 2012). 
The absence of any decrease of cluster frequency between 10 and 200 Myr (see Fig. \ref{fig4}), 
argues against the presence of dissolving clusters in our sample. We therefore assume that the observed
cluster response to gas expulsion is over by an age of 10 Myr and neglect the influence of residual
gas expulsion.

\subsection{Stellar evolution}

Stellar evolution reduces the masses of star clusters by about 45\% during a Hubble time if stars
follow a Kroupa (2001) IMF. Low-mass clusters with masses $M<10^4$ M$_\odot$ will therefore fall below
our adopted lower mass limit of $M=5000$ M$_\odot$ even without any dynamical mass loss. In order to model
stellar evolution, we assume for the mass lost by stellar evolution:
\begin{equation}
\frac{d M}{dt} = -\left( M_C - M_{Dyn}(t) \right) \frac{d \mu_{ev}}{dt}
\end{equation}
where $M_C$ is the initial cluster mass, $M_{Dyn}(t)$ is the accumulated mass lost by the dynamical
effects described below and $\mu_{ev}$ is calculated as described in \citet{letal10}, assuming a
metallicity of $Z=0.0080$. Apart from mass loss, stellar evolution also causes the clusters to expand
and very extended clusters could suffer additional dynamical mass loss due to this expansion and
the related tidal overflow. However we neglected the influence of this process since it is most effective in the
first few 100 Myr of cluster evolution, where we see little evidence for cluster dissolution (see Sec. 3). 

\subsection{Two-body relaxation and external tidal fields}

The effects of two-body relaxation and a spherical external tidal field were modeled according
to the results of \citet{bm03}, who performed simulations of multi-mass clusters moving through spherically
symmetric, isothermal galaxies. According to \citet{bm03}, the lifetime $t_{Dis|Rel}$ of a star cluster moving through
an external galaxy with circular velocity $V_C$ on an orbit with distance $R$ from the centre of the galaxy 
is given by
\begin{equation}
 \frac{t_{Dis|Rel}}{\mbox[Gyr]} = k \left( \frac{N}{ln(0.02 \, N)} \right)^x
  \!\! \frac{R}{\mbox{[kpc]}}
  \left( \frac{V_C}{220 \mbox{km/s}} \right)^{-1} \!\!\!\! .
\label{gtime}
\end{equation}
Here $N$ is the initial number of cluster stars, which can be calculated from the initial cluster
mass and the mean mass of the cluster stars as $N=M_C/\!\!<\!m\!>$. A \citet{k01} IMF between mass
limits of 0.1 and 100 $M_\odot$ has $<\!m\!>=0.64$ M$_\odot$. $x$ and $k$ 
are constants describing the dissolution process and are given by $x = 0.75$ and $k = 1.91 \cdot 10^{-3}$ \citep{bm03}.
Kinematic studies of various population tracer populations show that the LMC has near-solid body
rotation in its inner parts \citep[][e.g.]{metal88,hetal91}.
\citet{vdm02} found that the LMC shows solid body rotation out to a radius of about 4 kpc and an approximately flat rotation 
curve for larger radii. The rotation velocity of the LMC is however rather uncertain and depends on the tracer population that is used.
\citet{om07} find values of $V_C=61$ km/sec for carbon stars, $V_C=80$ km/sec for H~I gas  and $V_C=107$~km/sec
for red supergiants and speculate that the differences could be a sign of the ongoing tidal perturbation of the LMC.
If we assume a maximum circular rotation speed of the LMC of 80 km/sec, similar to what \citet{om07} found
for the rotation velocity of the H~I gas, eq. \ref{gtime} becomes
\begin{equation}
\nonumber \frac{t_{Dis|Rel}}{\mbox{[Gyr]}} = 3.92 \left( \frac{M_C}{10^4 M_\odot} \right)^{0.75} 
\label{tdisrel}
\end{equation}
for radii $R< 4$ kpc. We do not consider larger radii since observed LMC clusters mostly have $R<4$ kpc.
Note that, as a result of the solid body rotation, $t_{Dis|Rel}$ does not depend on galactocentric distance.

We finally note that \cite{gb08b} have shown that compact clusters dissolve faster than the tidally filling clusters
simulated by \citet{bm03} due to their smaller relaxation times. However we neglect this effect since it
only becomes important for $r_h/r_J<0.05$, where $r_h$ is the half-mass radius and $r_J$ is the Jacobi radius
of the cluster. A cluster starting with
an initial mass of $10^4$~M$_\odot$ at a typical distance of 1 kpc from the centre of the LMC has a Jacobi radius $r_J=20$ pc,
so this effect would only be important if clusters start with radii $r_h \ll 1$ pc. 

\subsection{Disc shocks}

Clusters passing through galactic discs experience tidal heating
due to the difference in acceleration for stars in different parts of the cluster \citep{o72}.
The dissolution time against disc shocking is given by (Binney \& Tremaine 1987, eq. 7-72):
\begin{equation}
 t_{Dis|Shock} = \frac{T_\psi \sigma_*^2 V_z^2}{8 \bar{z^2} \bar{g_z^2}}
\label{tshock1}
\end{equation}
where $T_\psi$ is the orbital time of the cluster, $\sigma_*$ the 1D velocity dispersion of stars in the cluster,
$V_z$ the velocity with which the cluster passes the disc, $\bar{z^2}= r_h^2/3$ the root-mean square z distance
of stars in the cluster and $g_z$ the gravitational acceleration of stars due to the disc. Using the virial theorem, we find
\begin{equation}
 \frac{3}{2} M_C \sigma^2_* = \frac{\eta G M_C^2}{r_h} 
\end{equation}
where $\eta \approx 0.4$ for most density profiles. For a typical half-mass radius of $r_h=4$ pc, this gives
\begin{equation}
\frac{\sigma_*}{\mbox{km/sec}}=1.7 \sqrt{\frac{M_C}{10^4 \mbox{M}_\odot}} \;\; .
\end{equation}
The LMC is classified as a SBm galaxy \citep{dvf72}
and contains a relatively thick disc with an out-of-plane axial ratio of $\sim 0.3$ or larger \citep{vdm02}. The velocity dispersion of
stars perpendicular to the LMC disc depends on their age and increases for older stars, similar to the situation in the Milky Way
\citep{vdm09}. We adopt a velocity dispersion of $V_z=10$ km/sec, which is intermediate between the velocity dispersion found for
red supergiants and that of the H I gas \citep{vdm09}.  
If we also assume a circular velocity of $V=20$ km/sec at a distance of R=1 kpc from the centre of the LMC, we 
find $T_\psi=312$ Myr.
For an infinitely thin disc with radial scale length of $R_D=1.42$ kpc \citep{wn01} and a total mass in stars and gas of
$M_D=3.2 \cdot 10^9$ M$_\odot$ \citep{vdm02}, the acceleration in z direction is $g_z= 4.3$ pc/Myr$^2$ at $R=1$~kpc. Inserting
all the values in eq. \ref{tshock1} gives a dissolution time of roughly 110 Myr for a $10^4$ M$_\odot$ cluster.
Eq. \ref{tshock1} is however only valid for impulsive encounters in which the time scale for shocking is much smaller than
the orbital time for stars in the cluster. Slower encounters have a reduced effect since stars with orbital times $t_{orb} \ll t_{shock}$
can adiabatically adjust to the shocking. \citet{go97} give various forms for the adiabatic correction factor. If we use their
eq. 11, then
\begin{equation}
A(x)=\left(1+x^2/4\right)^{-3/2}
\end{equation}
where $x$ is the ratio of the angular velocity of stars in the cluster to the crossing timescale of the cluster through
the disc and is given by 
\begin{equation}
x=\frac{\sigma_*}{r_h} \frac{2 z_0}{V_z}
\end{equation}
For a vertical scale height of $z_0=270$ pc \citep{vdm02} and a cluster mass of $10^4$ M$_\odot$, $x \approx 20$, and
$A(x) \approx 1/1500$. Disc shocks therefore seem to have a negligible influence on the evolution of star
clusters in the LMC. Even if we set $V_z=50$ km/sec, corresponding to a cluster moving on a highly inclined orbit
through the disc and assume a smaller scale height of $z_0=100$ pc in the past, the lifetime of a $10^4$ M$_\odot$ 
cluster against disc shocks is still larger than that against two-body relaxation. We therefore neglect the influence 
of disc shocks on the cluster evolution.

\subsection{GMC encounters}

Similar to galactic discs, passing giant molecular clouds create transient tidal forces on a star cluster, which increases the 
internal energy of a cluster and leads to cluster dissolution \citep{s58}. Wielen (1985) found that giant molecular clouds dominate the
dissolution of star clusters in the solar neighborhood.  

The most thorough calculation of the influence of GMC encounters has been given by Gieles et al. (2006), who showed that 
considering only the total 
energy gain can overestimate the effect of GMC encounters since much of the energy gained by the cluster is carried 
away by escapers and that considering the mass loss of a cluster leads to a better description of the dissolution process.
Gieles et al. (2006) also showed that GMCs cannot destroy clusters with more than
$\sim 10^4$ M$_\odot$ in a single disruptive encounter if one takes the finite size of the GMCs into account. For 
the cumulative effect of many distant encounters, they derived the following expression for the lifetime of a star
cluster of mass $M_C$ and half-mass radius $r_h$: 
\begin{equation}
t_{dis|GMC} = \left( \frac{3 \sigma_{cn} \eta}{8 \pi^{3/2} g f G} \frac{r_h^2}{\bar{r}^2} \right) \left( \frac{M_C}{r_h^3} \right)
  \frac{1}{\Sigma_n \rho_n}
\end{equation} 
where $\rho_n=M_{GMC} \; n_{GMC}$ is the mass density of GMCs in a galaxy, $\Sigma_n$ the typical surface density of a
single GMC, $f$ is the ratio of relative mass loss to the relative energy gain of a cluster during an encounter, 
$g$ a numerical correction factor for close encounters,
$\sigma_{cn}$ is the typical relative velocity between a star cluster and a GMC, 
$\bar{r}$ the root-mean square radius
of the cluster, and $\eta$ a constant which depends on the density profile of the cluster. A King (1966) model with
$W_0$=5.0 has $\eta \approx 0.4$ and $(r_h^2/\bar{r}^2)=0.50$. The typical velocity dispersion of stars in the
LMC is $\sigma=20$ km/sec \citep{vdm02}. Fig. 12 in Gieles et al. shows that for a typical cluster mass of $10^4$ M$_\odot$
and a typical GMC mass of $10^5$ M$_\odot$, the correction factor $g$ is slightly larger than unity, so we adopt $g=1.5$.
We also assume $f=0.25$ as found by 
Gieles et al. through direct $N$-body simulations and an average cluster radius of $r_h=4$ pc. 
The mass density $\rho_n$ and the surface density $\Sigma_n$ of the GMCs are taken from the results of the NANTEN LMC 
molecular cloud survey \citep{fetal08}.
Fukui et al. found 270 molecular clouds with masses $M_{CO}>2 \cdot 10^4$ M$_\odot$ within a survey area of 30 deg$^2$.
Inside 2~kpc from the center of the LMC, they found a surface mass density of GMCs of 2 M$_\odot$/pc$^2$. Assuming
a disc scale height of 180 pc \citep{petal01} for the gas and that all GMCs are located within $\pm 1$ scale height of the
plane of the disc, gives a space density $\rho_n=0.0056$ M$_\odot$/pc$^3$. 
\citet{hetal10} found an average surface density of $\Sigma_n = 50$ M$_\odot/$pc$^2$ for individual GMCs in the LMC.
Inserting all values, we find for the lifetime of star clusters in the LMC:
\begin{equation}
\frac{t_{dis|GMC}}{[Gyr]} = 93.0 \left( \frac{M_C}{10^4 M_\odot} \right) 
\end{equation}
Unlike in the Milky Way, GMC encounters do not seem to be a dominant dissolution process for star clusters in the LMC,
the main reason being their lower space density as a result of the large disc scale height and their smaller
surface density ($\Sigma_n = 50$ M$_\odot/$pc$^2$ for clouds in the LMC vs. $\Sigma_n = 170$ M$_\odot/$pc$^2$
in the Milky Way, see Solomon et al. 1987). $N$-body simulations by \citet{w00} suggest that the large scale 
height of the LMC disc could be a result of the interaction with the Milky Way. Hence the importance of GMC encounters
might have been higher in the past when the LMC disc had a smaller scale height. It is however unlikely that the lifetime
would drop below a Hubble time even for the smallest clusters in our sample, and we therefore also neglect GMC encounters.

We are therefore left with two-body relaxation driven evaporation as the only dissolution process which seems to have a
significant influence on the LMC cluster system. In order to model relaxation driven evaporation, we assume that the
dynamical mass loss is linear in time, so the dynamically lost mass is given by:
\begin{equation}
\frac{d M}{dt} = -\left( M_C-M_{ev}(t) \right)/t_{Dis|Rel} \;\; ,
\end{equation}
where $M_{ev}(t)$ is the total mass lost by stellar evolution up to time $t$.

We assume that clusters form with the same rate as the field stars and use the star formation history of LMC field stars
as derived by \citet{hz09} to form clusters. We assume that clusters form with a power-law mass function
$d N(m) \sim m^{-\alpha} dm$ and adjust the exponent such that the mass function of clusters in the youngest age bin 
in Fig.\ \ref{fig5} is fitted.
We also assume that clusters follow an exponential density distribution in the LMC with scale length $R=1.42$ kpc,
similar to what is seen for the field stars, and that clusters move on circular orbits.
Clusters are then evolved according to the description presented in this section up to the
present time. The mass and age distribution of the surviving clusters with $M_C > 5000$ M$_\odot$ and absolute luminosities
$M_V>-3.5$ mag is then compared with the observational data for LMC clusters and we discuss the results in the next section. 
\begin{figure*}
\begin{minipage}{8.9cm}
\includegraphics[width=88mm]{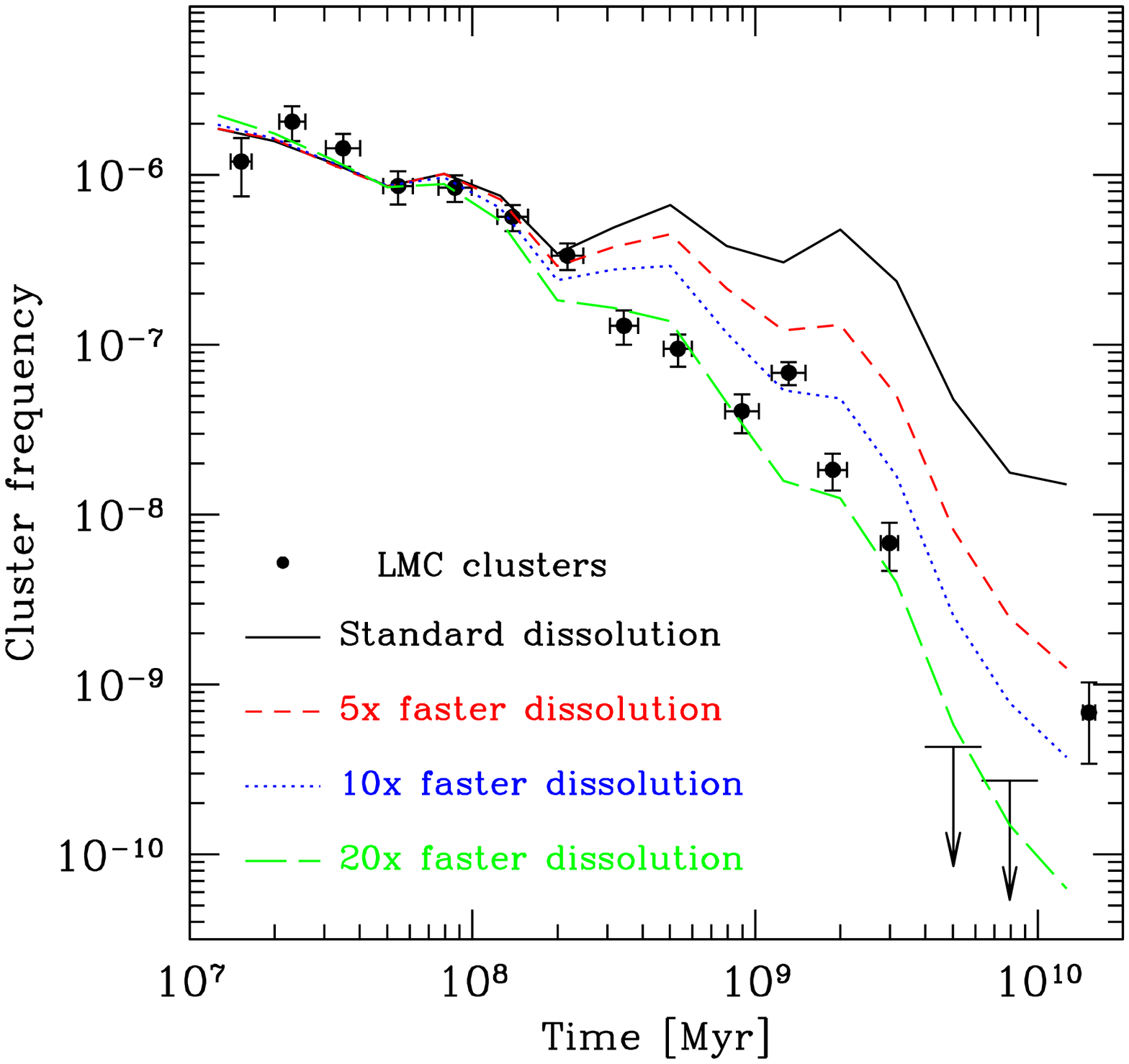}
\end{minipage}
\hfill
\begin{minipage}{8.1cm}
\includegraphics[width=81mm]{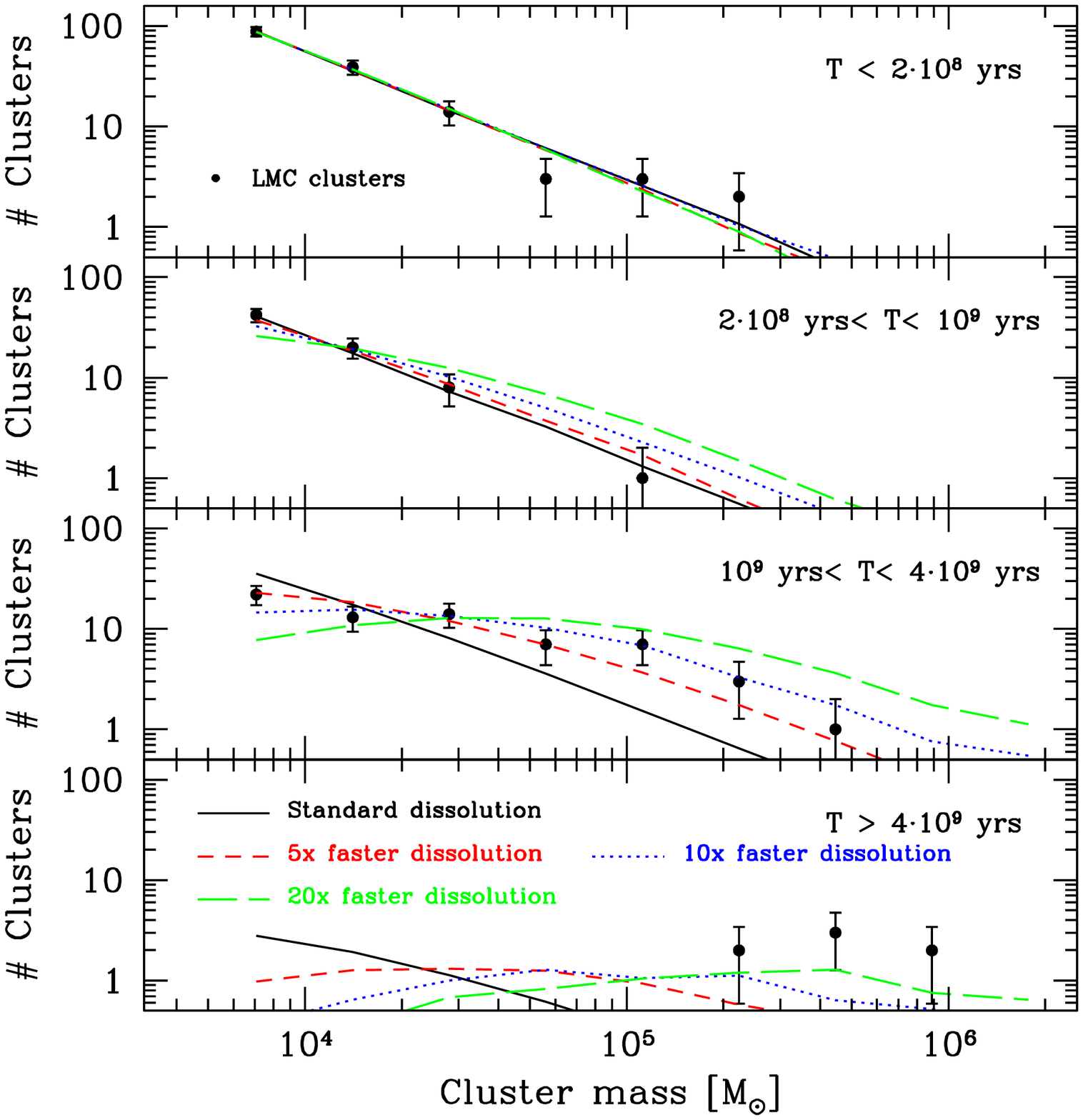}
\end{minipage}
\caption{Frequency of clusters as a function of time (left panel) and mass distribution of star clusters for different age bins
(right panel). Black dots in both panels depict the observed distribution of LMC clusters. The solid lines in both panels show
the theoretical prediction if clusters form with the same star formation rate as the field stars and the cluster
dissolution due to relaxation is applied as described in sec.\ \ref{sec:cldiss}. Red (short dashed), blue (dotted) and green (long dashed)
lines show predicted distributions if the cluster lifetimes are reduced by factors of 5, 10 and 20 respectively. In order to fit the data,
cluster dissolution has to be significantly faster than predicted by theory.}
\label{fig:varalpha}
\end{figure*}

\section{Comparison with theory}
\label{sec:cth}

The left panel of Fig.\ \ref{fig:varalpha} shows a comparison between the predicted and observed cluster frequency as a function of time, 
assuming that star clusters form with a power-law mass function 
$N(m) \sim m^{-\alpha}$ and with a rate similar to the star formation rate determined by \citet{hz09} for LMC field stars. 
Clusters are distributed in the LMC following an exponential density distribution up to a maximum radius of $R=4$ kpc, similar to the
largest distances from the centre of the LMC for clusters in our sample.
The black, solid line depicts the expected cluster frequency distribution if dissolution mechanisms are applied to star clusters 
exactly as described in the previous section. It can be seen that the frequency of the predicted clusters decreases significantly
slower with time than the cluster frequency of the observed clusters for ages $T>200$ Myr, and the black solid line
predicts about a factor 10 more clusters than observed for ages of several Gyr. If due to incompleteness, several hundred unstudied 
clusters with ages $1 < T < 4$ Gyr would still exist in the LMC. While new clusters in this age range are still being found \citep[e.g.][]{p11},
it seems rather unlikely that so many clusters are missing from our sample.

It is therefore much more likely that the theoretical estimates from sec.\ \ref{sec:cldiss} underestimate the
true rate of
cluster dissolution. The red (short dashed), blue (dotted) and green (long dashed) lines show the predicted cluster distribution if we
reduce the cluster lifetimes by factors of 5, 10 and 20 respectively. It can be seen that lifetimes need to be reduced by about a factor 
of at least 10 to 20 to give an agreement between predicted and observed cluster frequency.

The right panel of Fig.\ \ref{fig:varalpha} depicts the mass distribution of star clusters as a function of their age.
If we apply the various dissolution mechanisms as described in the previous section (solid lines),  
the mass distribution of star clusters up to 1 Gyr is in 
good agreement with the observational data since the predicted mass function can be fitted by the same slope  
for all ages, similar to what was found for the observed cluster distribution in sec.~3.
The reason is that in this case the turnover carved into the cluster mass function  remains at masses lower
than our mass-cut, and is therefore not seen in our cluster sample. As a result, at masses higher
than 5000 M$_\odot$, the slope of the cluster mass function is preserved during the first Gyr.
For ages larger than $10^9$ yrs, dissolution of clusters more massive than 5000 M$_\odot$ becomes important 
and the slope starts to flatten at the low-mass end of our sample. For clusters that are more than 4 Gyr old and standard dissolution, the
power-law mass function develops a turn-over at a few thousand solar masses. The location of this turn-over is 
at significantly smaller masses than the observed one, since the observed one is around $2\cdot 10^5$ M$_\odot$. 
In order to bring
the mass function of predicted clusters in the oldest
age bin into better agreement with observations a reduction in the lifetimes by about a factor 20 is needed 
(see green, long-dashed curve). 
In the age bin from $10^9$ yrs and $4 \cdot 10^9$ yrs,
lifetimes reduced by a factor 20 produce a mass function that also has a turnover, which is not present in the observed
data. A reduction by a factor of 5 to 10 is here
in much better agreement with the observations.  

The comparison with observations indicates that cluster lifetimes are shorter by about a factor 10 to 20 than given by the 
theoretical estimates of sec.\ \ref{sec:cldiss}. Given that most of the results in that section were derived based on
$N$-body simulations, which are to a large degree free of underlying assumptions, such a reduction seems to be outside the
uncertainties of the theoretical estimates. It also does not seem possible to achieve such a reduction in lifetime 
by changing the adopted LMC parameters, like the rotational velocity or the velocity dispersion, within the uncertainties 
quoted in the literature. Either the mode
of star formation and the properties of LMC clusters were drastically different at ages of 200 Myr and larger, or 
our knowledge of star formation in the LMC and the star clusters which were formed at these ages is still incomplete.
For example, if the star formation rate at ages $T>1$ Gyr
was smaller than estimated by \citet{hz09}, the difference in frequency between the predicted and observed curves
would be smaller and the reduction in lifetime necessary to fit the observed cluster frequencies would be smaller.
In addition, the globular cluster system of the LMC might have formed with a log-normal distribution instead of a 
power-law \citep{v98, pg07}, alleviating the need to drastically modify cluster lifetimes in order to predict the cluster mass distribution
in the oldest age bin.

\section{Conclusions}

We have compiled a new catalogue of ages and masses of LMC star clusters by combining results from four different surveys. Our catalogue
covers the whole LMC and contains data for 307 clusters with masses $M>5000$ M$_\odot$, ages $T>10^7$~yrs and absolute
magnitudes $M_V<-3.5$. We find no significant influence of cluster dissolution for clusters younger than about 200 Myr, 
since both the ratio 
of the number of clusters divided by the absolute star formation rate as well as the mass function of clusters is independent
of time for these clusters. If residual gas expulsion is an important dissolution mechanism for star clusters, its influence
must be restricted to the first 10 Myr of cluster evolution or low-mass clusters with masses $M<5000$ M$_\odot$.
Young star clusters in the LMC form with a power-law mass function with slope $\alpha=2.3$. If we extrapolate this mass function
down to $10^2$~M$_\odot$, then about 15\% of all stars in the LMC form in bound star clusters that survive for at least 10 Myr.

For ages larger than $T=200$ Myr, the cluster frequency starts to drop and the ratio of cluster frequency to star formation rate
is about a factor 40 smaller at $T=4$ Gyr than what it was at $T=200$ Myr. In addition, the mass function of clusters flattens 
for clusters older than $T=1$ Gyr. The number of missing clusters in our catalogue needed to explain this flattening seems to be 
too large to be explained by incompleteness and we therefore conclude that most of the flattening is due to cluster dissolution.
The amount of cluster dissolution necessary to fit the observed cluster distribution is about a factor 10 higher than predicted 
by theory, indicating either that the effectiveness of the considered processes was significantly underestimated or that
older star cluster formed with a mass distribution significantly different from their younger counterparts.
 
\section*{Acknowledgments}

HB is supported by the Australian Research Council through Future Fellowship grant FT0991052.
EKG wishes to acknowledge support from the Sonderforschungsbereich "The Milky Way System" (SFB 881)
of the German Research Foundation (DFG), especially via subproject B5.
PA acknowledges funding by the National Natural Science Foundation of China (NSFC, grant number 11073001). 
GP acknowledges support from the Max-Planck-Institut f\"ur 
Radioastronomie (Bonn) in the form of a Research Fellowship.

\label{lastpage}

\begin{table*}
\caption{Properties of massive LMC clusters. Sources for the data are: 1: Glatt et al. (2010),
2: Pietrzynski \& Udalski (2000), 3: Popescu et al. (2012), 4: Additional clusters from the literature.
Absolute luminosities for the clusters from Pietrzynski \& Udalski (2000) are taken from Hunter et al. (2003).}
\begin{tabular}{lr@{\,\,}r@{\,\,}rr@{\,\,}r@{\,\,}rrrcccl}
\hline
Name &  \multicolumn{3}{c}{RA} & \multicolumn{3}{c}{DEC} & $M_V$ & \multicolumn{1}{c}{$\log$ age} & \multicolumn{1}{c}{$\Delta \log$ age} 
 & $\log$ M  & Data & Alternative \\ 
 & \multicolumn{3}{c}{[J2000]} & \multicolumn{3}{c}{[J2000]} & & \multicolumn{1}{c}{[yr]} &\multicolumn{1}{c}{[yr]}&[M$_\odot$]&Source&Name\\ 
\hline
NGC1466 & 3 & 44 &  32.9 & -71 & 40 &  13.0 & -7.59 & 10.10 &  0.01 &  5.31 & 4 &  \\ 
NGC1651 & 4 & 37 &  31.1 & -70 & 35 &   2.0 & -7.16 &  9.30 &  0.09 &  5.24 & 4 &  \\ 
NGC1644 & 4 & 37 &  39.0 & -66 & 11 &  58.0 & -5.61 &  9.19 &  0.05 &  4.32 & 4 &  \\ 
NGC1652 & 4 & 38 &  22.0 & -68 & 40 &  21.0 & -5.37 &  9.23 &  0.05 &  4.29 & 4 &  \\ 
NGC1841 & 4 & 45 &  23.9 & -83 & 59 &  56.0 & -7.80 & 10.09 &  0.01 &  5.12 & 4 &  \\ 
LW52 & 4 & 45 &  46.0 & -71 & 35 &  40.9 & -5.30 &  8.70 &  0.20 &  3.83 & 1 &  \\ 
SL37 & 4 & 46 &  46.0 & -72 & 23 &  40.9 & -4.99 &  8.70 &  0.20 &  3.71 & 1 &  \\ 
NGC1695 & 4 & 47 &  43.0 & -69 & 22 &  26.0 & -6.24 &  8.00 &  0.20 &  3.73 & 1 &  \\ 
NGC1698 & 4 & 49 &   4.0 & -69 & 6 &  52.9 & -7.09 &  8.00 &  0.20 &  4.07 & 1 &  \\ 
HS35 & 4 & 49 &  33.0 & -69 & 43 &  40.0 & -4.93 &  9.00 &  0.18 &  3.89 & 2 & KMHK124 \\ 
NGC1704 & 4 & 49 &  55.0 & -69 & 45 &  19.1 & -7.09 &  7.50 &  0.20 &  3.76 & 1 &  \\ 
HS37 & 4 & 50 &  28.9 & -68 & 42 &  42.1 & -4.65 &  9.25 &  0.23 &  4.02 & 2 & KMHK139 \\ 
NGC1711 & 4 & 50 &  37.3 & -69 & 59 &   4.0 & -8.88 &  7.70 &  0.05 &  4.21 & 4 &  \\ 
SL58 & 4 & 50 &  59.0 & -69 & 38 &  13.0 & -6.14 &  8.37 &  0.15 &  3.95 & 2 & KMHK153 \\ 
SL66 & 4 & 51 &  56.5 & -70 & 23 &  25.1 & -5.50 &  9.23 &  0.15 &  4.34 & 2 & KMHK180 \\ 
NGC1718 & 4 & 52 &  25.6 & -67 & 3 &   6.0 & -7.14 &  9.30 &  0.30 &  5.10 & 4 &  \\ 
SL75 & 4 & 52 &  55.7 & -68 & 55 &  11.6 & -6.49 &  8.20 &  0.30 &  3.96 & 2 & KMHK199 \\ 
KMHK208 & 4 & 53 &   7.2 & -69 & 2 &  36.0 & -6.18 &  8.58 &  0.28 &  4.10 & 2 &  \\ 
SL76 & 4 & 53 &   9.0 & -68 & 12 &  41.0 & -6.71 &  8.10 &  0.40 &  3.98 & 1 &  \\ 
NGC1732 & 4 & 53 &  11.8 & -68 & 39 &   1.1 & -6.72 &  7.84 &  0.17 &  3.83 & 2 &  \\ 
KMHK229 & 4 & 53 &  51.0 & -69 & 34 &  19.0 & -5.04 &  9.05 &  0.31 &  3.97 & 2 &  \\ 
NGC1751 & 4 & 54 &  12.0 & -69 & 48 &  25.0 & -6.83 &  9.15 &  0.05 &  4.60 & 4 &  \\ 
NGC1754 & 4 & 54 &  18.9 & -70 & 26 &  31.0 & -7.36 & 10.19 &  0.06 &  5.39 & 4 &  \\ 
NGC1735 & 4 & 54 &  19.0 & -67 & 6 &   1.1 & -7.79 &  7.70 &  0.20 &  4.17 & 1 &  \\ 
BSDL210 & 4 & 54 &  37.0 & -68 & 57 &  33.1 & -5.78 &  8.40 &  0.40 &  3.82 & 1 &  \\ 
BRHT43b & 4 & 54 &  44.0 & -68 & 57 &  45.0 & -6.26 &  8.30 &  0.60 &  3.94 & 1 &  \\ 
NGC1756 & 4 & 54 &  50.0 & -69 & 14 &  17.0 & -6.26 &  8.40 &  0.05 &  4.02 & 4 &  \\ 
SL106 & 4 & 55 &   5.0 & -69 & 40 &  26.0 & -7.18 &  7.70 &  0.20 &  3.93 & 1 &  \\ 
NGC1755 & 4 & 55 &  13.0 & -68 & 12 &  16.9 & -9.11 &  7.40 &  0.20 &  4.50 & 1 &  \\ 
SL105 & 4 & 55 &  23.0 & -68 & 32 &  29.0 & -6.45 &  8.00 &  0.20 &  3.81 & 1 &  \\ 
HS65 & 4 & 55 &  31.7 & -68 & 52 &  58.8 & -5.02 &  8.97 &  0.09 &  3.91 & 2 & H88-31 \\ 
KMHK292 & 4 & 55 &  34.0 & -69 & 26 &  53.2 & -7.78 &  7.20 &  0.20 &  3.84 & 1 &  \\ 
KMHK300 & 4 & 55 &  39.0 & -70 & 32 &  48.0 & -4.68 &  9.06 &  0.28 &  3.83 & 2 &  \\ 
NGC1777 & 4 & 55 &  48.9 & -74 & 17 &   3.0 & -7.38 &  9.08 &  0.15 &  4.75 & 4 &  \\ 
SL114 & 4 & 56 &   7.9 & -69 & 14 &  47.6 & -7.27 &  7.35 &  0.10 &  3.73 & 2 & KMHK305 \\ 
SL117 & 4 & 56 &  22.0 & -68 & 58 &   0.1 & -6.90 &  8.10 &  0.40 &  4.06 & 1 &  \\ 
SL116 & 4 & 56 &  25.1 & -68 & 48 &  13.9 & -6.88 &  7.68 &  0.03 &  3.80 & 2 & KMHK315 \\ 
NGC1767 & 4 & 56 &  26.0 & -69 & 24 &  11.9 & -8.41 &  7.30 &  0.40 &  4.15 & 1 &  \\ 
SL124e & 4 & 56 &  31.0 & -69 & 58 &  54.0 & -5.50 &  8.73 &  0.17 &  3.93 & 2 & KMHK324e \\ 
HD32228 & 4 & 56 &  34.0 & -66 & 28 &  25.0 & -7.77 &  7.30 &  0.20 &  3.89 & 1 &  \\ 
BSDL285 & 4 & 56 &  37.0 & -68 & 14 &  12.8 & -5.83 &  8.50 &  0.40 &  3.91 & 1 &  \\ 
SL119 & 4 & 56 &  38.1 & -68 & 9 &  53.3 & -6.16 &  8.86 &  0.08 &  4.28 & 2 & KMHK316 \\ 
NGC1772 & 4 & 56 &  52.0 & -69 & 33 &  22.0 & -7.76 &  7.60 &  0.20 &  4.11 & 1 &  \\ 
KMHK355 & 4 & 57 &  22.9 & -70 & 28 &  58.6 & -3.81 &  9.27 &  0.18 &  3.70 & 2 &  \\ 
SL134 & 4 & 57 &  30.0 & -68 & 21 &  47.0 & -6.79 &  7.57 &  0.09 &  3.70 & 2 & KMHK349 \\ 
SL136 & 4 & 57 &  30.6 & -69 & 3 &   6.5 & -5.80 &  9.19 &  0.11 &  4.39 & 2 & KMHK352 \\ 
NGC1782 & 4 & 57 &  51.0 & -69 & 23 &  35.2 & -8.30 &  7.20 &  0.20 &  4.05 & 1 &  \\ 
NGC1774 & 4 & 58 &   5.0 & -67 & 14 &  31.9 & -8.15 &  7.70 &  0.20 &  4.32 & 1 &  \\ 
BSDL365 & 4 & 58 &  40.0 & -70 & 31 &  39.0 & -5.36 &  8.70 &  0.60 &  3.85 & 1 &  \\ 
SL151 & 4 & 58 &  51.1 & -69 & 57 &  32.5 & -5.59 &  9.15 &  0.07 &  4.10 & 2 & KMHK388 \\ 
SL150 & 4 & 58 &  56.5 & -69 & 13 &   2.3 & -5.03 &  9.44 &  0.01 &  4.34 & 2 & KMHK386 \\ 
NGC1791 & 4 & 59 &   7.4 & -70 & 10 &   7.5 & -5.94 &  8.15 &  0.23 &  3.71 & 2 &  \\ 
NGC1786 & 4 & 59 &   7.9 & -67 & 44 &  45.0 & -8.41 & 10.18 &  0.01 &  5.57 & 4 &  \\ 
NGC1783 & 4 & 59 &   8.0 & -65 & 59 &  18.0 & -8.11 &  9.18 &  0.05 &  5.26 & 4 &  \\ 
SL153 & 4 & 59 &  19.0 & -66 & 19 &   7.0 & -6.92 &  8.20 &  0.40 &  4.14 & 1 &  \\ 
BRHT61a & 4 & 59 &  21.4 & -68 & 50 &  36.3 & -5.98 &  8.79 &  0.01 &  4.16 & 2 &  \\ 
NGC1793 & 4 & 59 &  37.0 & -69 & 33 &  27.0 & -6.40 &  8.00 &  0.40 &  3.79 & 1 &  \\ 
NGC1795 & 4 & 59 &  46.0 & -69 & 48 &   6.0 & -6.08 &  9.11 &  0.05 &  4.36 & 4 &  \\ 
NGC1801 & 5 & 0 &  35.6 & -69 & 36 &  50.1 & -6.91 &  8.45 &  0.11 &  4.31 & 2 &  \\ 
\hline
\end{tabular}
\end{table*}

\pagebreak
 
\setcounter{table}{1}
\begin{table*}
\caption{cont.}
\begin{tabular}{lr@{\,\,}r@{\,\,}rr@{\,\,}r@{\,\,}rrrcccl}
\hline
Name &  \multicolumn{3}{c}{RA} & \multicolumn{3}{c}{DEC} & $M_V$ & \multicolumn{1}{c}{$\log$ age} & \multicolumn{1}{c}{$\Delta 
\log$ age} & $\log$ M  & Data & Alternative \\ 
 & \multicolumn{3}{c}{[J2000]} & \multicolumn{3}{c}{[J2000]} & & \multicolumn{1}{c}{[yr]} &\multicolumn{1}{c}{[yr]}&[M$_\odot$]&Source&Name\\ 
\hline
HS87 & 5 & 0 &  41.0 & -69 & 20 &  31.0 & -4.56 &  9.25 &  0.05 &  3.98 & 3 &  \\ 
SL168 & 5 & 0 &  43.0 & -65 & 27 &  18.0 & -6.29 &  8.30 &  0.40 &  3.96 & 1 &  \\ 
HS85 & 5 & 0 &  51.0 & -67 & 48 &  14.0 & -4.71 &  9.02 &  0.04 &  3.82 & 2 & KMHK428 \\ 
NGC1804 & 5 & 1 &   4.0 & -69 & 4 &  57.0 & -6.83 &  7.80 &  0.40 &  3.85 & 1 &  \\ 
HS88 & 5 & 1 &   5.3 & -68 & 5 &   0.7 & -4.55 &  9.39 &  0.16 &  4.10 & 2 & KMHK436 \\ 
SL174 & 5 & 1 &  12.3 & -67 & 49 &   4.9 & -5.93 &  9.04 &  0.08 &  4.32 & 2 & KMHK439 \\ 
H88-93 & 5 & 1 &  29.4 & -67 & 38 &   0.7 & -3.72 &  9.48 &  0.37 &  3.85 & 2 &  \\ 
KMHK448 & 5 & 1 &  29.5 & -68 & 42 &  42.2 & -5.54 &  8.45 &  0.08 &  3.77 & 3 &  \\ 
SL180 & 5 & 1 &  37.0 & -69 & 2 &  18.0 & -5.54 &  9.20 &  0.10 &  4.32 & 3 &  \\ 
SL181 & 5 & 1 &  51.0 & -69 & 12 &  50.0 & -5.17 &  8.80 &  0.10 &  3.85 & 3 &  \\ 
NGC1806 & 5 & 2 &  11.0 & -67 & 59 &  17.0 & -7.50 &  9.18 &  0.05 &  5.01 & 4 &  \\ 
NGC1805 & 5 & 2 &  21.0 & -66 & 6 &  42.1 & -8.42 &  7.60 &  0.20 &  4.37 & 1 &  \\ 
NGC1815 & 5 & 2 &  27.0 & -70 & 37 &  14.9 & -6.57 &  7.80 &  0.20 &  3.75 & 1 &  \\ 
HS94 & 5 & 2 &  30.6 & -70 & 17 &  33.5 & -4.39 &  9.05 &  0.19 &  3.71 & 2 & KMHK468 \\ 
SL188 & 5 & 2 &  33.7 & -68 & 49 &  24.0 & -6.16 &  8.10 &  0.10 &  3.76 & 3 &  \\ 
SL191 & 5 & 3 &   6.1 & -69 & 2 &  12.0 & -7.00 &  8.10 &  0.05 &  4.10 & 3 &  \\ 
SL197 & 5 & 3 &  34.3 & -67 & 37 &  32.2 & -4.95 &  9.10 &  0.08 &  3.93 & 3 & KMHK482 \\ 
HS102 & 5 & 3 &  38.5 & -69 & 23 &  10.0 & -4.78 &  9.10 &  0.05 &  3.87 & 3 &  \\ 
NGC1818 & 5 & 4 &  13.8 & -66 & 26 &   2.0 & -9.61 &  7.40 &  0.20 &  4.13 & 4 &  \\ 
NGC1825 & 5 & 4 &  19.4 & -68 & 55 &  40.0 & -7.24 &  7.80 &  0.10 &  4.02 & 3 &  \\ 
NGC1828 & 5 & 4 &  21.0 & -69 & 23 &  17.0 & -6.69 &  8.40 &  0.10 &  4.18 & 3 &  \\ 
KMK88-8 & 5 & 4 &  30.0 & -69 & 9 &  21.0 & -5.25 &  8.60 &  0.10 &  3.75 & 3 & H88-106 \\ 
KMK88-7 & 5 & 4 &  31.0 & -69 & 21 &  19.0 & -5.16 &  9.00 &  0.10 &  3.99 & 3 & H88-108 \\ 
NGC1830 & 5 & 4 &  39.0 & -69 & 20 &  27.1 & -6.25 &  8.60 &  0.10 &  4.14 & 3 &  \\ 
NGC1837 & 5 & 4 &  53.0 & -70 & 42 &  55.1 & -7.77 &  7.30 &  0.20 &  3.89 & 1 &  \\ 
KMK88-11 & 5 & 5 &   4.1 & -68 & 54 &  35.8 & -5.75 &  9.05 &  0.15 &  4.25 & 3 & H88-117 \\ 
NGC1835 & 5 & 5 &   6.7 & -69 & 24 &  15.0 & -8.74 & 10.22 &  0.07 &  5.83 & 4 &  \\ 
NGC1834 & 5 & 5 &  12.0 & -69 & 12 &  27.0 & -7.12 &  8.10 &  0.15 &  4.15 & 3 &  \\ 
SL212 & 5 & 5 &  12.4 & -68 & 33 &  10.7 & -6.49 &  8.70 &  0.10 &  4.31 & 3 &  \\ 
HS112 & 5 & 5 &  33.3 & -69 & 6 &  59.2 & -5.39 &  8.95 &  0.04 &  4.04 & 2 &  \\ 
NGC1836 & 5 & 5 &  35.7 & -68 & 37 &  42.0 & -6.84 &  8.70 &  0.05 &  4.45 & 3 &  \\ 
SL224 & 5 & 5 &  43.3 & -70 & 19 &  30.1 & -5.49 &  8.50 &  0.42 &  3.77 & 2 & KMHK534 \\ 
HS111 & 5 & 5 &  44.9 & -68 & 30 &  23.9 & -4.39 &  9.10 &  0.10 &  3.71 & 3 &  \\ 
HS115 & 5 & 5 &  54.8 & -70 & 22 &  21.4 & -4.98 &  8.97 &  0.25 &  3.89 & 2 & KMHK537 \\ 
NGC1839 & 5 & 6 &   1.0 & -68 & 37 &  36.1 & -7.17 &  7.90 &  0.20 &  4.05 & 1 &  \\ 
NGC1831 & 5 & 6 &  17.4 & -64 & 55 &  11.0 & -8.41 &  8.50 &  0.30 &  4.81 & 4 &  \\ 
HS117 & 5 & 6 &  25.9 & -68 & 42 &  12.0 & -5.10 &  9.20 &  0.10 &  4.15 & 3 &  \\ 
SL228w & 5 & 6 &  28.0 & -66 & 54 &  24.1 & -6.49 &  8.20 &  0.20 &  3.96 & 1 &  \\ 
SL230 & 5 & 6 &  33.0 & -68 & 21 &  47.9 & -7.55 &  7.40 &  0.20 &  3.88 & 1 &  \\ 
SL234 & 5 & 6 &  53.0 & -68 & 43 &   7.0 & -6.31 &  7.95 &  0.20 &  3.73 & 1 &  \\ 
OGLE-LMC0114 & 5 & 6 &  56.0 & -69 & 25 &  48.0 & -3.88 &  9.25 &  0.10 &  3.71 & 3 &  \\ 
SL237 & 5 & 6 &  58.3 & -69 & 9 &   0.1 & -7.11 &  7.78 &  0.05 &  3.95 & 3 &  \\ 
NGC1847 & 5 & 7 &   7.7 & -68 & 58 &  17.0 & -10.41 &  7.42 &  0.30 &  4.97 & 4 &  \\ 
NGC1844 & 5 & 7 &  29.0 & -67 & 19 &  23.9 & -6.31 &  7.90 &  0.40 &  3.70 & 1 &  \\ 
NGC1846 & 5 & 7 &  35.0 & -67 & 27 &  39.0 & -7.82 &  9.17 &  0.05 &  5.10 & 4 &  \\ 
SL249 & 5 & 7 &  35.5 & -70 & 44 &  56.1 & -6.14 &  8.38 &  0.14 &  3.95 & 2 & KMHK562 \\ 
SL244 & 5 & 7 &  38.9 & -68 & 32 &  30.9 & -6.12 &  9.43 &  0.01 &  4.77 & 2 &  \\ 
SL250 & 5 & 7 &  51.0 & -69 & 26 &  10.6 & -5.83 &  9.00 &  0.10 &  4.25 & 3 &  \\ 
NGC1850 & 5 & 8 &  45.8 & -68 & 45 &  38.0 & -10.95 &  7.50 &  0.20 &  5.42 & 4 &  \\ 
BRHT5 & 5 & 8 &  54.6 & -68 & 45 &  16.8 & -7.28 &  8.20 &  0.10 &  4.28 & 3 & H88-159 \\ 
BSDL734 & 5 & 9 &  13.3 & -69 & 16 &  57.6 & -4.10 &  9.25 &  0.05 &  3.80 & 3 &  \\ 
SL268 & 5 & 9 &  14.8 & -69 & 35 &  16.0 & -7.02 &  9.20 &  0.30 &  4.92 & 3 &  \\ 
NGC1854 & 5 & 9 &  20.3 & -68 & 50 &  55.0 & -8.70 &  8.05 &  0.05 &  4.75 & 3 &  \\ 
NGC1852 & 5 & 9 &  23.0 & -67 & 46 &  42.0 & -6.49 &  9.12 &  0.05 &  4.51 & 4 &  \\ 
NGC1856 & 5 & 9 &  31.5 & -69 & 7 &  46.0 & -10.39 &  8.12 &  0.30 &  5.25 & 4 &  \\ 
NGC1849 & 5 & 9 &  34.0 & -66 & 18 &  59.0 & -5.85 &  8.30 &  0.20 &  3.78 & 1 &  \\ 
BSDL767 & 5 & 9 &  43.3 & -70 & 18 &  34.9 & -5.47 &  8.51 &  0.26 &  3.77 & 2 &  \\ 
HS141 & 5 & 9 &  49.3 & -69 & 5 &   3.1 & -5.01 &  9.00 &  0.10 &  3.92 & 3 &  \\ 
SL276 & 5 & 9 &  58.0 & -69 & 21 &  11.0 & -6.17 &  8.90 &  0.08 &  4.32 & 3 &  \\ 
OGLE-LMC0169 & 5 & 10 &   6.1 & -69 & 5 &  19.6 & -6.32 &  8.25 &  0.05 &  3.93 & 2 &  \\ 
SL282 & 5 & 10 &  10.9 & -70 & 22 &  33.3 & -4.52 &  9.13 &  0.10 &  3.71 & 2 & KMHK619 \\ 
SL278 & 5 & 10 &  16.0 & -68 & 29 &  30.8 & -5.53 &  8.40 &  0.40 &  3.72 & 1 &  \\ 
KMK88-32 & 5 & 10 &  20.0 & -68 & 52 &  45.0 & -5.58 &  8.50 &  0.10 &  3.81 & 3 & H88-178 \\ 
\hline
\end{tabular}
\end{table*}

\pagebreak
 
\setcounter{table}{1}
\begin{table*}
\caption{cont.}
\begin{tabular}{lr@{\,\,}r@{\,\,}rr@{\,\,}r@{\,\,}rrrcccl}
\hline
Name &  \multicolumn{3}{c}{RA} & \multicolumn{3}{c}{DEC} & $M_V$ & \multicolumn{1}{c}{$\log$ age} & \multicolumn{1}{c}{$\Delta 
\log$ age} & $\log$ M  & Data & Alternative \\ 
 & \multicolumn{3}{c}{[J2000]} & \multicolumn{3}{c}{[J2000]} & & \multicolumn{1}{c}{[yr]} &\multicolumn{1}{c}{[yr]}&[M$_\odot$]&Source&Name\\ 
\hline
NGC1861 & 5 & 10 &  21.9 & -70 & 46 &  44.2 & -5.97 &  8.79 &  0.11 &  4.16 & 2 &  \\ 
HS153 & 5 & 10 &  30.0 & -68 & 52 &  21.0 & -5.84 &  8.55 &  0.10 &  3.95 & 3 & BRHT48 \\ 
SL291 & 5 & 10 &  30.5 & -70 & 54 &  36.2 & -5.54 &  8.72 &  0.11 &  3.94 & 2 & KMHK626 \\ 
NGC1860 & 5 & 10 &  38.9 & -68 & 45 &  12.0 & -8.88 &  8.28 &  0.30 &  4.30 & 4 &  \\ 
SL288 & 5 & 10 &  39.4 & -69 & 2 &  28.6 & -6.94 &  7.65 &  0.05 &  3.81 & 3 &  \\ 
SL294 & 5 & 10 &  42.2 & -70 & 3 &  46.8 & -5.89 &  8.28 &  0.30 &  3.78 & 2 & KMHK627 \\ 
SL296 & 5 & 10 &  56.7 & -69 & 33 &  31.1 & -5.86 &  8.85 &  0.05 &  4.15 & 3 &  \\ 
NGC1859 & 5 & 11 &  31.0 & -65 & 14 &  57.1 & -6.31 &  8.10 &  0.20 &  3.82 & 1 &  \\ 
NGC1863 & 5 & 11 &  39.0 & -68 & 43 &  48.0 & -7.91 &  7.80 &  0.40 &  4.28 & 1 &  \\ 
SL304 & 5 & 12 &   1.0 & -69 & 12 &   1.0 & -6.76 &  8.40 &  0.10 &  4.21 & 3 &  \\ 
KMK88-38 & 5 & 12 &   9.3 & -68 & 54 &  40.6 & -4.40 &  9.10 &  0.05 &  3.71 & 3 & H88-206 \\ 
NGC1865 & 5 & 12 &  25.0 & -68 & 46 &  23.0 & -6.65 &  8.80 &  0.05 &  4.44 & 3 &  \\ 
BSDL880 & 5 & 12 &  27.6 & -69 & 33 &  13.8 & -4.68 &  9.10 &  0.05 &  3.83 & 3 &  \\ 
KMK88-40 & 5 & 12 &  34.0 & -69 & 17 &  11.0 & -7.66 &  7.20 &  0.60 &  3.79 & 1 &  \\ 
KMK88-42 & 5 & 12 &  50.2 & -68 & 51 &  51.2 & -6.34 &  8.20 &  0.05 &  3.91 & 3 & H88-216 \\ 
NGC1878 & 5 & 12 &  50.7 & -70 & 28 &  20.2 & -6.00 &  8.54 &  0.03 &  4.01 & 2 &  \\ 
HS177 & 5 & 13 &   3.5 & -69 & 3 &   1.7 & -4.48 &  9.25 &  0.05 &  3.95 & 3 &  \\ 
NGC1870 & 5 & 13 &  10.0 & -69 & 7 &   1.0 & -7.49 &  8.15 &  0.05 &  4.33 & 3 &  \\ 
NGC1872 & 5 & 13 &  11.0 & -69 & 18 &  43.0 & -8.05 &  8.70 &  0.10 &  4.93 & 3 &  \\ 
NGC1866 & 5 & 13 &  38.9 & -65 & 27 &  52.0 & -9.98 &  8.12 &  0.30 &  4.63 & 4 &  \\ 
BSDL946 & 5 & 13 &  57.5 & -68 & 42 &  52.1 & -5.67 &  9.13 &  0.06 &  4.17 & 2 &  \\ 
NGC1868 & 5 & 14 &  36.2 & -63 & 57 &  14.0 & -7.58 &  8.74 &  0.30 &  4.53 & 4 &  \\ 
HS190 & 5 & 14 &  47.0 & -69 & 27 &  22.0 & -5.44 &  9.44 &  0.35 &  4.51 & 2 &  \\ 
HS186 & 5 & 14 &  49.0 & -66 & 11 &   1.0 & -4.84 &  8.80 &  0.60 &  3.71 & 1 &  \\ 
HS191 & 5 & 14 &  51.0 & -69 & 25 &  39.5 & -4.58 &  8.95 &  0.10 &  3.71 & 3 &  \\ 
BSDL985 & 5 & 14 &  53.0 & -66 & 3 &  33.8 & -5.15 &  8.60 &  0.60 &  3.71 & 1 &  \\ 
NGC1885 & 5 & 15 &   6.0 & -68 & 58 &  45.0 & -7.32 &  8.30 &  0.03 &  4.37 & 3 &  \\ 
NGC1894 & 5 & 15 &  51.0 & -69 & 28 &   9.0 & -7.96 &  8.05 &  0.05 &  4.45 & 3 &  \\ 
NGC1887 & 5 & 16 &   4.0 & -66 & 19 &   9.1 & -6.13 &  8.10 &  0.20 &  3.75 & 1 &  \\ 
BSDL1102 & 5 & 16 &  37.4 & -70 & 12 &  38.9 & -4.12 &  9.21 &  0.11 &  3.77 & 2 &  \\ 
SL352 & 5 & 16 &  41.6 & -70 & 32 &  27.0 & -5.80 &  8.61 &  0.03 &  3.97 & 2 & KMHK715 \\ 
NGC1898 & 5 & 16 &  42.4 & -69 & 39 &  25.0 & -7.82 & 10.15 &  0.07 &  5.88 & 4 &  \\ 
SL349 & 5 & 16 &  55.1 & -68 & 52 &  36.2 & -5.38 &  8.85 &  0.05 &  3.96 & 3 & BRHT33 \\ 
H1 & 5 & 17 &   8.4 & -68 & 52 &  27.0 & -6.82 &  9.20 &  0.10 &  4.84 & 3 & SL353, BRHT33 \\ 
NGC1903 & 5 & 17 &  22.0 & -69 & 20 &  17.0 & -9.28 &  8.08 &  0.03 &  5.00 & 3 &  \\ 
SL357 & 5 & 17 &  27.0 & -69 & 22 &  35.0 & -6.59 &  9.20 &  0.08 &  4.75 & 3 & BRHT9 \\ 
SL358 & 5 & 17 &  34.0 & -69 & 30 &  51.0 & -6.03 &  8.70 &  0.05 &  4.12 & 3 &  \\ 
HS211 & 5 & 17 &  38.0 & -68 & 58 &  30.4 & -5.01 &  8.70 &  0.10 &  3.71 & 3 &  \\ 
H2 & 5 & 17 &  49.2 & -69 & 38 &  38.6 & -7.06 &  9.25 &  0.10 &  4.98 & 3 & SL363 \\ 
HS213 & 5 & 17 &  56.2 & -69 & 34 &  56.2 & -5.79 &  8.30 &  0.07 &  3.76 & 3 &  \\ 
BRHT10 & 5 & 18 &  10.7 & -69 & 32 &  25.4 & -5.87 &  8.20 &  0.10 &  3.72 & 3 & H88-264 \\ 
NGC1902 & 5 & 18 &  17.0 & -66 & 37 &  37.9 & -7.24 &  8.00 &  0.20 &  4.13 & 1 &  \\ 
NGC1913 & 5 & 18 &  19.1 & -69 & 32 &  13.7 & -7.55 &  7.58 &  0.03 &  4.01 & 3 &  \\ 
NGC1916 & 5 & 18 &  37.5 & -69 & 24 &  25.0 & -8.93 & 10.20 &  0.09 &  5.79 & 4 &  \\ 
H88-269 & 5 & 18 &  41.7 & -69 & 4 &  46.5 & -4.98 &  9.00 &  0.10 &  3.91 & 3 &  \\ 
HS223A & 5 & 18 &  52.4 & -69 & 22 &  15.2 & -5.15 &  9.10 &  0.05 &  4.02 & 3 &  \\ 
NGC1917 & 5 & 19 &   2.0 & -69 & 0 &   4.0 & -6.17 &  9.11 &  0.05 &  4.42 & 4 &  \\ 
HS227 & 5 & 19 &   4.0 & -69 & 48 &  39.0 & -5.16 &  9.00 &  0.03 &  3.99 & 3 &  \\ 
NGC1921 & 5 & 19 &  23.8 & -69 & 47 &  16.2 & -5.95 &  8.30 &  0.10 &  3.82 & 3 &  \\ 
SL385 & 5 & 19 &  26.3 & -69 & 32 &  25.2 & -6.63 &  8.50 &  0.05 &  4.23 & 3 & BRHT35 \\ 
SL387 & 5 & 19 &  33.7 & -69 & 32 &  31.7 & -6.41 &  9.00 &  0.10 &  4.49 & 3 & BRHT35 \\ 
NGC1922 & 5 & 19 &  50.0 & -69 & 30 &   1.0 & -7.74 &  7.35 &  0.10 &  3.92 & 3 &  \\ 
SL390 & 5 & 19 &  54.3 & -68 & 57 &  50.4 & -5.92 &  9.20 &  0.03 &  4.48 & 3 &  \\ 
SL397 & 5 & 20 &  12.0 & -68 & 54 &  15.1 & -6.67 &  7.80 &  0.20 &  3.79 & 1 &  \\ 
SL402 & 5 & 20 &  23.6 & -69 & 35 &   6.3 & -6.53 &  8.50 &  0.07 &  4.19 & 3 &  \\ 
H88-281 & 5 & 20 &  26.0 & -69 & 15 &   8.8 & -5.81 &  8.27 &  0.22 &  3.74 & 2 &  \\ 
NGC1926 & 5 & 20 &  35.0 & -69 & 31 &  27.8 & -7.76 &  8.20 &  0.10 &  4.47 & 3 &  \\ 
NGC1928 & 5 & 20 &  57.5 & -69 & 28 &  41.6 & -6.83 &  9.23 &  0.05 &  4.87 & 3 &  \\ 
BSDL1334 & 5 & 21 &  14.0 & -68 & 47 &   0.0 & -3.72 &  9.48 &  0.41 &  3.86 & 2 &  \\ 
NGC1938 & 5 & 21 &  25.4 & -69 & 56 &  23.3 & -5.90 &  8.75 &  0.05 &  4.10 & 3 &  \\ 
NGC1939 & 5 & 21 &  27.1 & -69 & 57 &   1.3 & -7.43 &  9.20 &  0.10 &  5.08 & 3 &  \\ 
SL410 & 5 & 21 &  45.0 & -65 & 13 &  55.9 & -6.36 &  8.05 &  0.20 &  3.81 & 1 &  \\ 
SL418 & 5 & 21 &  49.5 & -69 & 39 &   6.9 & -6.47 &  8.40 &  0.20 &  4.10 & 3 &  \\ 
\hline
\end{tabular}
\end{table*}

\pagebreak
 
\setcounter{table}{1}
\begin{table*}
\caption{cont.}
\begin{tabular}{lr@{\,\,}r@{\,\,}rr@{\,\,}r@{\,\,}rrrcccl}
\hline
Name &  \multicolumn{3}{c}{RA} & \multicolumn{3}{c}{DEC} & $M_V$ & \multicolumn{1}{c}{$\log$ age} & \multicolumn{1}{c}{$\Delta 
\log$ age} & $\log$ M  & Data & Alternative \\ 
 & \multicolumn{3}{c}{[J2000]} & \multicolumn{3}{c}{[J2000]} & & \multicolumn{1}{c}{[yr]} &\multicolumn{1}{c}{[yr]}&[M$_\odot$]&Source&Name\\ 
\hline
NGC1944 & 5 & 21 &  57.0 & -72 & 29 &  39.0 & -6.66 &  7.84 &  0.05 &  3.81 & 4 &  \\ 
SL419 & 5 & 22 &   3.4 & -69 & 15 &  18.3 & -6.39 &  8.55 &  0.05 &  4.17 & 3 &  \\ 
SL423 & 5 & 22 &  13.3 & -69 & 30 &  48.8 & -6.30 &  8.40 &  0.10 &  4.03 & 3 &  \\ 
SL425 & 5 & 22 &  25.0 & -68 & 47 &   6.0 & -6.21 &  8.20 &  0.40 &  3.85 & 1 &  \\ 
NGC1932 & 5 & 22 &  26.0 & -66 & 9 &   9.0 & -6.85 &  8.10 &  0.20 &  4.04 & 1 &  \\ 
NGC1943 & 5 & 22 &  29.5 & -70 & 9 &  14.9 & -7.25 &  8.35 &  0.05 &  4.38 & 3 &  \\ 
NGC1940 & 5 & 22 &  43.0 & -67 & 11 &  10.0 & -6.50 &  8.00 &  0.20 &  3.83 & 1 &  \\ 
HS259 & 5 & 22 &  45.7 & -69 & 50 &  49.5 & -7.18 &  8.14 &  0.02 &  4.20 & 2 &  \\ 
SL434 & 5 & 23 &  25.6 & -69 & 1 &  18.2 & -5.07 &  8.70 &  0.20 &  3.74 & 2 &  \\ 
SL453 & 5 & 25 &   1.9 & -69 & 26 &   7.4 & -7.01 &  8.50 &  0.05 &  4.38 & 3 &  \\ 
SL461 & 5 & 25 &  19.0 & -71 & 48 &  11.2 & -5.05 &  8.70 &  0.40 &  3.73 & 1 &  \\ 
NGC1953 & 5 & 25 &  26.0 & -68 & 50 &  17.9 & -6.87 &  7.90 &  0.20 &  3.93 & 1 &  \\ 
NGC1951 & 5 & 26 &   4.0 & -66 & 35 &  49.9 & -7.82 &  7.70 &  0.20 &  4.18 & 1 &  \\ 
BSDL1657 & 5 & 26 &   5.0 & -67 & 10 &  57.0 & -6.33 &  8.00 &  0.20 &  3.77 & 1 &  \\ 
KMHK898 & 5 & 26 &  24.0 & -68 & 2 &  48.1 & -5.59 &  9.12 &  0.03 &  4.16 & 2 &  \\ 
HS301A & 5 & 26 &  38.0 & -71 & 58 &  50.2 & -6.46 &  7.90 &  0.20 &  3.76 & 1 &  \\ 
NGC1967 & 5 & 26 &  43.0 & -69 & 6 &   5.0 & -7.62 &  7.20 &  0.20 &  3.78 & 1 &  \\ 
NGC1987 & 5 & 27 &  17.0 & -70 & 44 &   8.0 & -6.76 &  9.03 &  0.05 &  4.65 & 4 &  \\ 
SL482 & 5 & 27 &  17.0 & -66 & 22 &   7.0 & -7.47 &  7.60 &  0.20 &  3.99 & 1 &  \\ 
NGC2000 & 5 & 27 &  29.0 & -71 & 52 &  48.0 & -6.62 &  8.00 &  0.20 &  3.88 & 1 &  \\ 
NGC1984 & 5 & 27 &  40.0 & -69 & 8 &   3.1 & -9.11 &  7.80 &  0.20 &  4.76 & 1 &  \\ 
SL492 & 5 & 27 &  43.0 & -68 & 59 &   8.2 & -7.47 &  7.70 &  0.20 &  4.04 & 1 &  \\ 
SL495 & 5 & 28 &   3.0 & -68 & 48 &  42.1 & -7.59 &  7.40 &  0.20 &  3.89 & 1 &  \\ 
NGC1994 & 5 & 28 &  21.0 & -69 & 8 &  30.1 & -7.02 &  7.70 &  0.40 &  3.86 & 1 &  \\ 
HS314 & 5 & 28 &  26.0 & -68 & 58 &  55.9 & -7.30 &  7.30 &  0.40 &  3.70 & 1 &  \\ 
KMHK945 & 5 & 28 &  27.4 & -68 & 38 &  59.4 & -5.27 &  9.08 &  0.25 &  4.06 & 2 &  \\ 
SL498 & 5 & 28 &  34.0 & -67 & 13 &  30.0 & -7.15 &  7.45 &  0.12 &  3.75 & 2 & KMHK943 \\ 
HODGE14 & 5 & 28 &  39.3 & -73 & 37 &  49.0 & -5.95 &  9.26 &  0.10 &  4.33 & 4 &  \\ 
NGC1978 & 5 & 28 &  45.0 & -66 & 14 &  10.0 & -7.80 &  9.30 &  0.05 &  5.33 & 4 &  \\ 
HS319 & 5 & 28 &  47.6 & -68 & 59 &   1.4 & -6.02 &  8.32 &  0.41 &  3.86 & 2 & BRHT52a \\ 
SL503 & 5 & 29 &   0.2 & -68 & 25 &   8.0 & -5.47 &  8.56 &  0.42 &  3.81 & 2 & KMHK952 \\ 
SL502 & 5 & 29 &  10.0 & -66 & 35 &  29.0 & -8.37 &  7.60 &  0.20 &  4.35 & 1 &  \\ 
BSDL1938 & 5 & 29 &  19.0 & -69 & 0 &  20.9 & -7.16 &  7.60 &  0.20 &  3.87 & 1 &  \\ 
OGLE-LMC0531 & 5 & 30 &   2.1 & -69 & 31 &  36.2 & -4.24 &  9.30 &  0.05 &  3.91 & 3 &  \\ 
NGC2005 & 5 & 30 &  10.3 & -69 & 45 &   9.0 & -13.12 & 10.22 &  0.14 &  5.49 & 4 &  \\ 
NGC2002 & 5 & 30 &  21.0 & -66 & 53 &   2.0 & -8.57 &  7.10 &  0.20 &  4.05 & 1 &  \\ 
ESO86SC2 & 5 & 30 &  22.0 & -65 & 54 &  32.0 & -7.39 &  7.50 &  0.20 &  3.88 & 1 &  \\ 
BSDL2001 & 5 & 30 &  25.4 & -67 & 13 &  17.9 & -6.74 &  7.81 &  0.18 &  3.82 & 2 &  \\ 
NGC2010 & 5 & 30 &  35.0 & -70 & 49 &  11.0 & -6.78 &  8.19 &  0.05 &  4.07 & 4 &  \\ 
NGC2004 & 5 & 30 &  40.9 & -67 & 17 &   9.0 & -9.62 &  7.30 &  0.20 &  4.43 & 4 &  \\ 
NGC2003 & 5 & 30 &  53.0 & -66 & 27 &  59.0 & -7.78 &  7.40 &  0.20 &  3.97 & 1 &  \\ 
SL539 & 5 & 30 &  55.0 & -70 & 41 &  42.0 & -7.89 &  7.40 &  0.20 &  4.01 & 1 &  \\ 
NGC2009 & 5 & 30 &  58.0 & -69 & 11 &   3.1 & -7.60 &  7.50 &  0.20 &  3.96 & 1 &  \\ 
SL543 & 5 & 30 &  59.0 & -71 & 53 &  35.9 & -6.38 &  7.90 &  0.20 &  3.73 & 1 &  \\ 
KMHK1022 & 5 & 31 &   7.0 & -71 & 57 &  45.0 & -6.68 &  8.30 &  0.60 &  4.11 & 1 &  \\ 
SL538 & 5 & 31 &  17.0 & -66 & 57 &  28.1 & -7.92 &  7.40 &  0.20 &  4.03 & 1 &  \\ 
NGC2006 & 5 & 31 &  19.0 & -66 & 58 &  22.1 & -7.81 &  7.60 &  0.20 &  4.13 & 1 &  \\ 
NGC2019 & 5 & 31 &  56.6 & -70 & 9 &  33.0 & -12.97 & 10.25 &  0.08 &  5.68 & 4 &  \\ 
BSDL2180 & 5 & 32 &   1.0 & -66 & 50 &  57.1 & -5.01 &  8.70 &  0.60 &  3.71 & 1 &  \\ 
HS346 & 5 & 32 &   3.0 & -69 & 22 &  10.0 & -6.11 &  8.22 &  0.20 &  3.82 & 2 &  \\ 
SL558 & 5 & 32 &  11.9 & -69 & 29 &  41.1 & -5.66 &  8.40 &  0.10 &  3.77 & 3 &  \\ 
KMHK1047 & 5 & 32 &  18.0 & -68 & 52 &  28.0 & -6.59 &  7.74 &  0.19 &  3.72 & 2 &  \\ 
BRHT14b & 5 & 32 &  19.0 & -67 & 31 &  40.1 & -7.25 &  7.50 &  0.20 &  3.82 & 1 &  \\ 
NGC2011 & 5 & 32 &  19.0 & -67 & 31 &  16.0 & -7.66 &  7.40 &  0.20 &  3.92 & 1 &  \\ 
HODGE4 & 5 & 32 &  25.2 & -64 & 44 &  11.0 & -9.29 &  9.34 &  0.10 &  5.39 & 4 &  \\ 
NGC2025 & 5 & 32 &  33.0 & -71 & 43 &   0.8 & -7.36 &  7.95 &  0.20 &  4.15 & 1 &  \\ 
BSDL2300 & 5 & 33 &  19.0 & -68 & 53 &  32.0 & -5.86 &  9.47 &  0.03 &  4.70 & 2 &  \\ 
SL569 & 5 & 33 &  20.4 & -68 & 9 &   9.9 & -5.83 &  9.08 &  0.01 &  4.29 & 2 & KMHK1065 \\ 
NGC2031 & 5 & 33 &  41.1 & -70 & 59 &  13.0 & -8.98 &  8.20 &  0.10 &  5.13 & 4 &  \\ 
SL582 & 5 & 34 &  28.0 & -67 & 7 &  28.9 & -7.23 &  7.90 &  0.20 &  4.07 & 1 &  \\ 
HS358 & 5 & 34 &  35.0 & -66 & 3 &  56.2 & -6.05 &  8.30 &  0.20 &  3.86 & 1 &  \\ 
KMHK1098 & 5 & 34 &  40.0 & -67 & 30 &  13.0 & -6.95 &  7.50 &  0.20 &  3.70 & 1 &  \\ 
SL588 & 5 & 34 &  40.3 & -68 & 18 &  18.5 & -5.46 &  8.70 &  0.10 &  3.89 & 2 & KMHK1101 \\ 
\hline
\end{tabular}
\end{table*}

\pagebreak
 
\setcounter{table}{1}
\begin{table*}
\caption{cont.}
\begin{tabular}{lr@{\,\,}r@{\,\,}rr@{\,\,}r@{\,\,}rrrcccl}
\hline
Name &  \multicolumn{3}{c}{RA} & \multicolumn{3}{c}{DEC} & $M_V$ & \multicolumn{1}{c}{$\log$ age} & \multicolumn{1}{c}{$\Delta 
\log$ age} & $\log$ M  & Data & Alternative \\ 
 & \multicolumn{3}{c}{[J2000]} & \multicolumn{3}{c}{[J2000]} & & \multicolumn{1}{c}{[yr]} &\multicolumn{1}{c}{[yr]}&[M$_\odot$]&Source&Name\\ 
\hline
SL586 & 5 & 34 &  43.0 & -66 & 57 &  45.0 & -7.03 &  7.60 &  0.20 &  3.81 & 1 &  \\ 
HS359 & 5 & 34 &  45.5 & -69 & 23 &  18.4 & -5.98 &  9.02 &  0.04 &  4.33 & 2 &  \\ 
KMHK1112 & 5 & 35 &  16.6 & -68 & 52 &  13.9 & -3.77 &  9.38 &  0.11 &  3.78 & 2 &  \\ 
NGC2030 & 5 & 35 &  39.0 & -66 & 1 &  50.2 & -6.32 &  8.00 &  0.60 &  3.76 & 1 &  \\ 
HDE269828 & 5 & 36 &   0.4 & -69 & 11 &  50.5 & -7.97 &  7.18 &  0.09 &  3.90 & 2 &  \\ 
NGC2051 & 5 & 36 &   7.0 & -71 & 0 &  42.1 & -6.92 &  8.00 &  0.40 &  4.00 & 1 &  \\ 
LT-delta & 5 & 36 &  10.0 & -69 & 11 &  46.0 & -7.94 &  7.30 &  0.20 &  3.96 & 1 &  \\ 
NGC2041 & 5 & 36 &  28.0 & -66 & 59 &  26.2 & -8.52 &  7.40 &  0.40 &  4.27 & 1 &  \\ 
SL607 & 5 & 36 &  31.4 & -68 & 48 &  44.6 & -5.66 &  8.59 &  0.07 &  3.90 & 2 & KMHK1137 \\ 
NGC2056 & 5 & 36 &  34.0 & -70 & 40 &  18.0 & -6.73 &  8.40 &  0.05 &  4.20 & 4 &  \\ 
NGC2053 & 5 & 37 &  40.5 & -67 & 24 &  51.4 & -6.46 &  8.09 &  0.27 &  3.87 & 2 &  \\ 
IC2146 & 5 & 37 &  46.0 & -74 & 46 &  58.0 & -6.09 &  9.19 &  0.05 &  4.51 & 4 &  \\ 
M-OB1 & 5 & 38 &  16.0 & -69 & 3 &  58.0 & -7.76 &  7.40 &  0.20 &  3.96 & 1 &  \\ 
SL629 & 5 & 38 &  22.0 & -68 & 46 &  50.7 & -4.45 &  9.31 &  0.02 &  4.00 & 2 & KMHK1169 \\ 
SL628 & 5 & 38 &  30.0 & -67 & 19 &  57.0 & -6.44 &  8.30 &  0.15 &  4.02 & 2 & KMHK1166 \\ 
BSDL2652 & 5 & 38 &  31.1 & -68 & 8 &  23.0 & -3.93 &  9.24 &  0.11 &  3.73 & 2 &  \\ 
KMHK1188 & 5 & 39 &  30.0 & -68 & 19 &  21.0 & -4.08 &  9.48 &  0.02 &  4.00 & 2 &  \\ 
NGC2091 & 5 & 40 &  58.0 & -69 & 26 &  11.0 & -6.95 &  7.70 &  0.40 &  3.84 & 1 &  \\ 
NGC2088 & 5 & 40 &  58.0 & -68 & 27 &  52.9 & -6.32 &  7.90 &  0.20 &  3.71 & 1 &  \\ 
BSDL2794 & 5 & 41 &  10.0 & -69 & 10 &  44.0 & -5.55 &  9.07 &  0.24 &  4.18 & 2 &  \\ 
NGC2093 & 5 & 41 &  49.0 & -68 & 55 &  15.0 & -7.22 &  7.48 &  0.15 &  3.80 & 2 &  \\ 
KMHK1244 & 5 & 42 &   0.0 & -67 & 20 &  34.1 & -6.79 &  7.90 &  0.20 &  3.90 & 1 &  \\ 
NGC2100 & 5 & 42 &   8.6 & -69 & 12 &  44.0 & -9.93 &  7.20 &  0.20 &  4.48 & 4 &  \\ 
NGC2096 & 5 & 42 &  16.0 & -68 & 27 &  29.9 & -7.51 &  7.50 &  0.20 &  3.93 & 1 &  \\ 
SL663 & 5 & 42 &  28.8 & -65 & 21 &  44.0 & -7.58 &  9.51 &  0.06 &  5.23 & 4 &  \\ 
NGC2098 & 5 & 42 &  29.0 & -68 & 16 &  28.9 & -7.78 &  7.60 &  0.20 &  4.11 & 1 &  \\ 
NGC2108 & 5 & 43 &  56.0 & -69 & 10 &  50.0 & -6.18 &  9.01 &  0.05 &  4.40 & 4 &  \\ 
NGC2105 & 5 & 44 &  19.0 & -66 & 55 &   4.1 & -6.23 &  8.20 &  0.20 &  3.86 & 1 &  \\ 
NGC2109 & 5 & 44 &  22.0 & -68 & 32 &  48.1 & -6.58 &  8.20 &  0.40 &  4.00 & 1 &  \\ 
BM32 & 5 & 44 &  49.0 & -67 & 19 &  43.0 & -6.67 &  7.90 &  0.20 &  3.85 & 1 &  \\ 
SL714 & 5 & 47 &  16.0 & -66 & 52 &  59.2 & -5.28 &  8.60 &  0.40 &  3.76 & 1 &  \\ 
NGC2117 & 5 & 47 &  46.0 & -67 & 27 &   2.9 & -6.70 &  7.80 &  0.20 &  3.80 & 1 &  \\ 
NGC2121 & 5 & 48 &  11.6 & -71 & 28 &  51.0 & -7.85 &  9.51 &  0.06 &  5.69 & 4 &  \\ 
HODGE7 & 5 & 50 &   3.0 & -67 & 43 &   5.0 & -6.24 &  9.17 &  0.05 &  4.47 & 4 &  \\ 
SL748 & 5 & 50 &  15.0 & -70 & 25 &  40.1 & -5.85 &  8.40 &  0.40 &  3.85 & 1 &  \\ 
BM112 & 5 & 50 &  21.0 & -68 & 39 &  27.0 & -5.64 &  8.40 &  0.60 &  3.77 & 1 &  \\ 
LW318 & 5 & 50 &  43.0 & -65 & 18 &  14.0 & -6.48 &  8.10 &  0.20 &  3.89 & 1 &  \\ 
NGC2127 & 5 & 51 &  22.0 & -69 & 21 &  38.9 & -6.90 &  7.90 &  0.20 &  3.94 & 1 &  \\ 
NGC2133 & 5 & 51 &  28.0 & -71 & 10 &  28.9 & -6.30 &  8.20 &  0.20 &  3.89 & 1 &  \\ 
NGC2123 & 5 & 51 &  43.0 & -65 & 19 &  17.0 & -6.01 &  8.20 &  0.20 &  3.77 & 1 &  \\ 
NGC2134 & 5 & 51 &  55.0 & -71 & 5 &  51.0 & -7.71 &  7.40 &  0.40 &  3.94 & 1 &  \\ 
NGC2130 & 5 & 52 &  22.0 & -67 & 20 &   2.0 & -6.31 &  7.90 &  0.40 &  3.70 & 1 &  \\ 
NGC2136 & 5 & 52 &  58.0 & -69 & 29 &  36.0 & -8.62 &  8.00 &  0.10 &  4.45 & 4 &  \\ 
HS445 & 5 & 53 &  49.0 & -67 & 23 &   8.9 & -6.77 &  8.30 &  0.20 &  4.15 & 1 &  \\ 
NGC2145 & 5 & 54 &  22.0 & -70 & 54 &   4.0 & -6.27 &  8.30 &  0.20 &  3.95 & 1 &  \\ 
NGC2157 & 5 & 57 &  32.4 & -69 & 11 &  49.0 & -9.09 &  7.60 &  0.20 &  4.31 & 4 &  \\ 
NGC2154 & 5 & 57 &  38.0 & -67 & 15 &  43.0 & -6.71 &  9.15 &  0.05 &  4.57 & 4 &  \\ 
NGC2156 & 5 & 57 &  45.0 & -68 & 27 &  38.2 & -7.17 &  7.90 &  0.20 &  4.05 & 1 &  \\ 
NGC2153 & 5 & 57 &  51.2 & -66 & 23 &  58.0 & -5.58 &  9.11 &  0.14 &  3.97 & 4 &  \\ 
NGC2159 & 5 & 57 &  57.0 & -68 & 37 &  26.0 & -7.10 &  8.00 &  0.20 &  4.08 & 1 &  \\ 
NGC2173 & 5 & 57 &  58.5 & -72 & 58 &  40.0 & -7.33 &  9.33 &  0.08 &  5.06 & 4 &  \\ 
NGC2160 & 5 & 58 &  11.0 & -68 & 17 &  24.0 & -6.30 &  8.00 &  0.20 &  3.75 & 1 &  \\ 
NGC2155 & 5 & 58 &  33.3 & -65 & 28 &  35.0 & -6.68 &  9.51 &  0.06 &  4.90 & 4 &  \\ 
NGC2164 & 5 & 58 &  55.9 & -68 & 31 &   0.0 & -8.61 &  7.70 &  0.20 &  4.12 & 4 &  \\ 
NGC2172 & 6 & 0 &   4.0 & -68 & 38 &  12.8 & -6.80 &  7.90 &  0.20 &  3.90 & 1 &  \\ 
NGC2162 & 6 & 0 &  30.4 & -63 & 43 &  19.0 & -6.33 &  9.11 &  0.14 &  4.40 & 4 &  \\ 
NGC2190 & 6 & 1 &   2.0 & -74 & 43 &  33.0 & -5.56 &  9.04 &  0.05 &  4.17 & 4 &  \\ 
SL822 & 6 & 2 &   5.0 & -68 & 20 &   8.2 & -5.45 &  8.80 &  0.60 &  3.96 & 1 &  \\ 
NGC2203 & 6 & 4 &  42.0 & -75 & 26 &  16.0 & -7.21 &  9.26 &  0.05 &  5.05 & 4 &  \\ 
NGC2193 & 6 & 6 &  17.0 & -65 & 5 &  54.0 & -6.23 &  9.34 &  0.10 &  4.42 & 4 &  \\ 
SL842 & 6 & 8 &  14.9 & -62 & 59 &  15.0 & -5.34 &  9.30 &  0.09 &  4.08 & 4 &  \\ 
NGC2209 & 6 & 8 &  34.8 & -73 & 50 &  12.0 & -7.90 &  8.98 &  0.19 &  5.03 & 4 &  \\ 
NGC2213 & 6 & 10 &  42.2 & -71 & 31 &  46.0 & -6.48 &  9.20 &  0.11 &  4.56 & 4 &  \\ 
\hline
\end{tabular}
\end{table*}

\pagebreak
 
\setcounter{table}{1}
\begin{table*}
\caption{cont.}
\begin{tabular}{lr@{\,\,}r@{\,\,}rr@{\,\,}r@{\,\,}rrrcccl}
\hline
Name &  \multicolumn{3}{c}{RA} & \multicolumn{3}{c}{DEC} & $M_V$ & \multicolumn{1}{c}{$\log$ age} & \multicolumn{1}{c}{$\Delta 
\log$ age} & $\log$ M  & Data & Alternative \\ 
 & \multicolumn{3}{c}{[J2000]} & \multicolumn{3}{c}{[J2000]} & & \multicolumn{1}{c}{[yr]} &\multicolumn{1}{c}{[yr]}&[M$_\odot$]&Source&Name\\ 
\hline
SL855 & 6 & 10 &  53.7 & -65 & 2 &  30.0 & -5.01 &  9.13 &  0.30 &  4.58 & 4 &  \\ 
NGC2210 & 6 & 11 &  31.5 & -69 & 7 &  17.0 & -8.00 & 10.20 &  0.01 &  5.48 & 4 &  \\ 
NGC2214 & 6 & 12 &  55.8 & -68 & 15 &  38.0 & -8.41 &  7.60 &  0.20 &  4.50 & 4 &  \\ 
LW431 & 6 & 13 &  27.0 & -70 & 41 &  42.0 & -4.83 &  9.18 &  0.05 &  3.94 & 4 &  \\ 
HODGE11 & 6 & 14 &  22.3 & -69 & 50 &  50.0 & -7.50 & 10.18 &  0.01 &  5.63 & 4 &  \\ 
NGC2231 & 6 & 20 &  42.7 & -67 & 31 &  10.0 & -7.03 &  9.18 &  0.11 &  5.18 & 4 &  \\ 
NGC2241 & 6 & 22 &  53.0 & -68 & 55 &  30.0 & -5.25 &  9.28 &  0.05 &  4.29 & 4 &  \\ 
NGC2249 & 6 & 25 &  49.8 & -68 & 55 &  13.0 & -6.73 &  8.82 &  0.30 &  4.35 & 4 &  \\ 
NGC2257 & 6 & 30 &  12.1 & -64 & 19 &  42.0 & -7.06 & 10.20 &  0.10 &  5.41 & 4 &  \\ 
\hline
\end{tabular}
\end{table*}

\end{document}